\documentstyle[12pt,english]{report}
\def\intqp3{\int{d^3q\over (2\pi)^3 q_0 }  }
 \def\wlog#1{} 
\catcode`\@=11
 
\def\({\relax\ifmmode[\else$[$\nobreak\hskip.3em\fi}
\def\){\relax\ifmmode]\else\nobreak\hskip.2em$]$\fi}
 
\def\gappr{\mathpalette\under@rel{>\approx}}
\def\lappr{\mathpalette\under@rel{<\approx}}
\def\gsim{\mathpalette\under@rel{>\sim}}
\def\lsim{\mathpalette\under@rel{<\sim}}
\def\under@rel#1#2{\under@@rel#1#2}
 
\def\under@@rel#1#2#3{\mathrel{\mathop{#1#2}\limits_{#1#3}}}
 
\def\under@@rel#1#2#3{\mathrel{\vcenter{\hbox{$%
  \lower3.8pt\hbox{$#1#2$}\atop{\raise1.8pt\hbox{$#1#3$}}%
  $}}}}

\def\parenbar{\mathpalette\p@renb@r}
\def\p@renb@r#1#2{\vbox{%
  \ifx#1\scriptscriptstyle \dimen@.7em\dimen@ii.2em\else
  \ifx#1\scriptstyle \dimen@.8em\dimen@ii.25em\else
  \dimen@1em\dimen@ii.4em\fi\fi \offinterlineskip
  \ialign{\hfill##\hfill\cr
    \vbox{\hrule width\dimen@ii}\cr
    \noalign{\vskip-.3ex}%
    \hbox to\dimen@{$\mathchar300\hfil\mathchar301$}\cr
    \noalign{\vskip-.3ex}%
    $#1#2$\cr}}}
 
\def\mppae@text{{Max-Planck-Institut f\"ur Physik}}
\def\mppwh@text{{Werner-Heisenberg-Institut}}
 
\def\mppaddresstext{Postfach 40 12 12, D-80712 M\"unchen\else
  P.O.Box 40 12 12, 80712 Munich (Fed.^^>Rep.^^>Germany)}
 
\def\mppaddress{\address{\mppae@text \nl -- \mppwh@text\space --\nl
  \case@language\mppaddresstext}}
 
\font\fourteenssb=cmssdc10 scaled \magstep2 
\font\seventeenssb=cmssdc10 scaled \magstep3 
 
\def\letter#1#2{\b@lett@r{26}%
  \centerline{\seventeenssb \uppercase\mppae@text}%
  \centerline{\fourteenssb \uppercase\mppwh@text}%
  \centerline{\strut#1}\vskip.5cm%
  \e@lett@r{\hss\vtop to5cm{\hsize55mm%
    \lftline{\strut}\eightrm  \setbaselineskip=12pt \vfil
    \lftline{F\"OHRINGER RING 6}\lftline{\tenrm D-80805 M\"UNCHEN}%
    \lftline{\case@language{TELEFON\else PHONE}: (089) 3 23 54
      \if!#2!\else - #2 \case@language{oder\else or} \fi-0}%
    \lftline{TELEGRAMM:}\lftline{PHYSIKPLANCK M\"UNCHEN}%
    \lftline{TELEX: 5 21 56 19 mppa d}%
    \lftline{TELEFAX: \MPPfax}%
    \lftline{POSTFACH 40 12 12}%
    \ifx\EARN\undefined\else\vskip5\p@\lftline{EARN/BITNET: \EARN
      @dmumpiwh}\fi \vfil}}}
 
\def\MPPfax{(089) 3 22 67 04}
 
\def\b@lett@r#1{\endpage \begingroup \doublespace \vglue-#1mm}
 
\def\e@lett@r#1#2{\skippagenum T\skipheadline T\skipfootline T%
  \line{\vtop to47mm{\lftline{\llap{\vbox to\z@{\vskip171\p@
      \hrule\@width7\p@\vss}\hskip57\p@}\strut}\vskip2mm\vfil
    \addressspacing \dimen@\baselineskip \dimen@ii-2.79ex%
    \advance\dimen@ii\dimen@ \baselineskip\dimen@\@minus\dimen@ii
    \let\nl\cr\use@nl \halign{##\hfil\crcr#2\crcr}\vfil}#1}%
  \vskip1cm\rtline{\thedate}\vskip1cm\@plus1cm\@minus.5cm\endgroup}
 
\let\addressspacing=\empty
 
\def\myname{Dr.\ Xxxx Xxxxxxxxxx\nl Physiker}
\def\myaddress{Xxxxxxx Stra\ss e  ??\nl
    \llap{D--8????\quad}M\"unchen\nl
    Tel:\ (089) \vtop{\hbox{?? ?? ?? (privat)}%
                      \hbox{3 23 08-??? (B\"uro)}}}
\def\myletter{\b@lett@r{26}\line{\let\nl\cr \use@nl \caps
  \vtop to25mm{\halign{\strut##\hfil\crcr\myname\crcr}\vfil}\hfil
  \vtop to25mm{\tenpoint
    \halign{\strut##\hfil\crcr\myaddress\crcr}\vfil}}%
  \e@lett@r\empty}
 
\def\firstpageoutput{\physoutput
  \global\output{\setbox\z@\box@cclv \deadcycles\z@}}
 
\def\veq{\afterassignment\v@eq \dimen@}
\def\v@eq{$$\vcenter to\dimen@{}$$}
 
\def\veqn{\afterassignment\v@eqn \dimen@}
\def\v@eqn{$$\vcenter to\dimen@{}\eqn$$}
 
\def\heq{\afterassignment\h@eq \dimen@}
\def\h@eq{$\hbox to\dimen@{}$ }
 
\def\wlog{\immediate\write\m@ne} 
\catcode`\@=12 

\outer\def\pthnum#1{\errmessage{***** \string\pthnum\space is no longer
         supported, use \string\mppnum\space instead}}

\newenvironment{eqn}
{\begin{equation}\begin{array}}{\end{array}\end{equation}{}}

\def\EARN{HNS}
\def\myname{Dr.\ Heinrich Saller\nl Physiker}
\def\myaddress{Reiserbergweg 3\nl
    \llap{D--82327\quad}Tutzing\nl
    Tel:\ \vtop{\hbox{(08158) 1647 (privat)}%
                      \hbox{(089) 3 18 93-310 (B\"uro)}}}
 
\def\3{\ss }  
 
\def\({\Bigl(}
\def\){\Bigr)}
\def\|{\Big|}
\def\o{\circ}
\def\x{\times}

\def\ox{\otimes}

\def\OX{\displaystyle\bigotimes}
\def\pl{\oplus}

\def\PL{\displaystyle \bigoplus}
\def\SUM{\displaystyle \sum}

\def\mid{\big\bracevert}

\def\then{~\Rightarrow~}

\def\subnoteq{\subset}
\def\and{\wedge}

\def\A{{\,{\rm A\kern-.55emA}}}
\def\C{{\,{\rm I\kern-.55emC}}}
\def\E{{\,{\rm I\kern-.2emE}}}
\def\H{{\,{\rm I\kern-.2emH}}}
\def\I{{\,{\rm I\kern-.2emI}}}
\def\K{{\,{\rm I\kern-.2emK}}}
\def\L{{\,{\rm I\kern-.2emL}}}
\def\M{{\,{\rm I\kern-.16emM}}}
\def\N{{\,{\rm I\kern-.16emN}}}
\def\Q{{\,{\rm I\kern-.5emQ}}}
\def\R{{\,{\rm I\kern-.2emR}}}
\def\S{{\,{\rm I\kern-.42emS}}}
\def\T{{\,{\rm I\kern-.37emT}}}
\def\Z{{\,{\rm Z\kern-.35emZ}}}
\def\rin{{\,\in\kern-.42em\in}}

\def\tr{{\,{\rm tr }\,}}
\def\det{\,{\rm det }\,}

\def\supnoteq{\supset}

\def\p{\partial}

\def\al{\alpha}  \def\be{\beta} 
\def\de{\delta}  \def\ep{\epsilon}  
\def\th{\theta}    \def\io{\iota}
\def\ka{\kappa}   \def\la{\lambda}   \def\si{\sigma}
   \def\om{\omega} 
\def\phi{\varphi}    
    \def\La{\Lambda}

\let\rvec=\vec        
\def\vec#1{\underline{\bf vec}_{#1}}

 \def\GL{{\bf GL}}  \def\SL{{\bf SL}}
\def\U{{\bf U}} \def\O{{\bf O}}   \def\SU{{\bf SU}} \def\SO{{\bf SO}}
  
\def \UL{{\bf UL}}  \def\D{{\bl D}}

\def\d#1{\check{#1}}
\def\angle#1{\langle#1\rangle}

\def\brack#1{\lbrack#1\rbrack}

\def\ul#1{\underline{#1}}

\def\ol#1{\overline{#1}}
\def\bl#1{{\bf #1}}
\def\cl#1{{\cal #1}}

\def\ro#1{{\rm #1}}

\def\dprod#1#2{\langle#1,#2\rangle}
\def\sprod#1#2{\langle#1|#2\rangle}
\def\com#1#2{\lbrack#1,#2\rbrack}

\def\acom#1#2{\{#1,#2\}}

\def\bra#1#2{\lbrack\!\lbrack#1,#2\rbrack\!\rbrack}

\def\map{\longrightarrow}
\def\lrmap{\leftrightarrow}
\def\mape{\longmapsto}

\begin{document}
\begin{titlepage}

\hfill MPI-PhT/96-6 (January 96)
\vskip3cm
\centerline{\bf THE ANALYSIS OF}
\centerline{\bf TIME-SPACE TRANSLATIONS}
\centerline{\bf IN QUANTUM FIELDS}
\vskip2cm
\centerline{Heinrich Saller\footnote{\scriptsize 
e-mail adress: hns@mppmu.mpg.de}}
\centerline{Max-Planck-Institut f\"ur Physik und Astrophysik}
\centerline{Werner-Heisenberg-Institut f\"ur Physik}
\centerline{M\"unchen}
\vskip5mm
\centerline{January 1996}

\vskip2cm

\centerline{\bf Abstract}

I discuss the
indefinite metrical structure of the time-space translations as realized
in the  indefinite inner products for relativistic quantum fields,
familiar in the example of quantum  gauge fields. 
The arising indefinite unitary
nondiagonalizable representations of the translations 
suggest as the positive  unitarity condition for the probability interpretable
positive definite asymptotic particle state space the 
requirement of a vanishing  nilpotent part
in the time-space translations realization. A trivial
Becchi-Rouet-Stora charge (classical gauge invariance) 
for the asymptotics in quantum gauge theories
can be interpreted as one special case of this general principle -
the asymptotic
projection to the eigenstates of the time-space translations.

\end{titlepage}

\advance\topmargin by -1.6cm

\newpage


\centerline{\bf Notational Preliminaries}
\vskip5mm

Throughout this paper
 a definite basis for the - apparently - threefold dimensional graduation 
 in physics is assumed:
$\hbar$ (Planck's action scale), 
$c$ (Einstein's velocity scale) and an unspecified
mass scale $\mu_0$. With such a basis
all masses and energy-momenta come as real numbers. 

Relativistic fields
are symbolized with boldface letters, e.g. $\bl \Phi(x)$, $\bl Z(x)$,
$\bl l(x)$, $\bl b(x)$ etc., their harmonic components with roman letters, e.g.
$\ro e,\ro U,\ro a,\ro b$ etc.

For Lie groups,  
$\U(n_+,n_-)$ and $\SU(n_+,n_-)$ with $n_++n_-=n$ stand for the unitary
and special unitary  groups.
$\O(n_+,n_-)$  and 
$\SO(n_+,n_-)$ denote the real orthogonal groups, 
$\SO^+(1,n)$ the orthochronous groups.
The notations 
$\GL(\C^n)$, $\SL(\C^n)$ and $\GL(\R^n)$, $\SL(\R^n)$ are used for
the complex and real general $n^2$-dimensional 
and special $(n^2-1)$-di\-men\-sio\-nal  groups.
If $\GL(\C^n)$ and 
$\SL(\C^n)$ are considered as  real Lie groups with dimension
$2n^2$ and $2(n^2-1)$ resp. and maximal compact groups $\U(n)$ and $\SU(n)$, 
they are denoted by $\GL(\C^n)_\R$ and $\SL(\C^n)_\R$. 

For  groups realized in endomorphisms (matrix groups) a more
individual notation  proves useful. The $\U(1)$ isomorphic phase group
for a $d$-dimensional complex space is  written  as $\U(1_d)$. If
$\U(1)$ is realized in $\SU(2)$ by 
${\scriptsize\pmatrix{e^{i\al}&0\cr 0&e^{-i\al}\cr}}$, 
the notation
$\U(1)_3$ will be used, in $\SU(2d)$ the notation $\U(1_d)_3$.
If $\U(1)$ comes in $\U(2)$  as  
${\scriptsize\pmatrix{e^{i\al}&0\cr 0&1\cr}}$, 
it will be called $\U(1)_+$ and correspondingly $\U(1)_-$ and
$\U(1_d)_\pm$.
Analogue notations will be used also for other groups, e.g.
$\SL(\C^n_2)_\R$ for
${\scriptsize\pmatrix{\SL(\C^n)_\R&0\cr 0&\SL(\C^n)_\R\cr}}$. 

The groups $\U(n_+,n_-)$ are the product of two normal subgroups, the 
phase group
and the special group $\U(1_n)\o\SU(n_+,n_-)$. 
Because of the cyclic group $\I_n=\{z\in\C\mid z^n=1\}$
as intersection of both normal subgroups $\U(1_n)\cap\SU(n_+,n_-)\cong\I_n$
the product is not direct for $n\ge2$.
The group $\GL(\C^n)_\R$ is the direct product 
$\GL(\C^n)_\R=\D(1_n)\x \UL(\C^n)_\R$ of the
normal subgroups $\D(1_n)$ 
(dilatations) and $\UL(\C^n)_\R=\U(1_n)\o\SL(\C^n)_\R$, the latter one
being the product of  the phase group
and the special linear group, not direct for $n\ge2$.

The Lie algebras for the groups will be denoted with corresponding small letters,
e.g. $\bl u(1_d)\cong\bl u(1)$ for $\U(1_d)$, 
$\bl{sl}(\C^n_d)_\R\cong\bl{sl}(\C^n)_\R$ for $\SL(\C^n_d)_\R$ etc.

\pagebreak 


\centerline{\bf INTRODUCTION}
\vskip1cm

Wigner's particle classification 
\cite{WIG} relies on the harmonic analysis of the
Poin\-ca\-r\'e group 
in terms of  $\U(1)$-characters for time-space translations, i.e.
positive unitary representations  $e^{ i xq}\in\U(1)$
with real  energies $q_0=\sqrt{m^2+\rvec q^2}$. 
The semidirect product Poincar\'e group $\SO^+(1,3)\x_s\M$ with the
ortho\-chro\-nous Lorentz group $\SO^+(1,3)$ and the 
Minkowski time-space translations
$\M\cong\R^4$ as action group for fields is reduced  for particles
to a direct product group $\O\x\R$ with a homogeneous compact
group $\O\subnoteq\SO^+(1,3)$ as the stability group for a 1-dimensional
translation group $\R$.

For time translations $\T\cong\R$,  spanned with the 
nontrivial mass $m^2=q^2>0$ of a particle, the stability group $\SO(3)$
describes the rotation degrees of freedom  of the rest frames which are
characterized by the energy-momenta $\ul q(m)=(m,0,0,0)$.
An associated  Sylvester decomposition
splits the Minkowski space
$\M\cong\T\pl\S^3$ into time and space translations $\S^3\cong\R^3$.
  
For lightlike momenta  $q^2=0$, $q\ne0$,  and massless fields the Minkowski
translations have to be Witt-decomposed 
$\M\cong\L_+\pl\S^2\pl \L_-$ into two 1-di\-men\-sio\-nal 
lightlike translation spaces
$\L_\pm\cong\R$ and  2-dimensional space translations $\S^2\cong\R^2$.
The stability group of  those time-space translations
frames which are 
determined by two independent lightlike vectors 
$\ul q(\mu_\pm)=\mu_\pm(1,0,0,\pm1)$  or - equivalently -
by one nontrivial timelike and one spacelike vector 
$\L_+\pl\L_-\cong\T\pl\S^1$ with
$\ul q(\mu)=(\mu,0,0,0)$ and $\ul q(\ka)=(0,0,0,\ka)$, 
is the circularity (helicity, polarization) group $\SO(2)$.

Collecting both cases, there arises the following scheme of  Minkowski
space decompositions with their particles relevant stability groups
\[
\SO^+(1,3)\hbox{ for }\M\begin{array}{l}
\supnoteq\left\{\begin{array}{c}
\SO(3)\hbox{ for }\T\pl\S^3\\
(m^2>0) \end{array}\right.\\
~~\\
\supnoteq\left\{\begin{array}{c}
\SO(2)\hbox{ for }\L_+\pl\S^2\pl\L_-\\
(m^2=0)\end{array}\right. \end{array}
\]

In the complex framework of quantum theory 
the Lorentz symmetry comes as the group
$\SL(\C^2)$, considered as real 6-dimensional 
Lie group and denoted by $\SL(\C^2)_\R$, 
with
the isomorphy $\SO^+(1,3)\cong \SL(\C^2)_\R/\I_2$ where $\I_2=\{\pm1\}$
is the sign group (real phases).  Even more: 
The special linear group $\SL(\C^2)_\R$ comes as normal subgroup of the 
real 7-dimensional phase Lorentz group
\[\UL(\C^2)_\R=\{\la\in\GL(\C^2)_\R\mid |\det\la|=1\}\]
The orthochronous Lorentz group is the manifold 
of the phase $\U(1_2)$-orbits in the phase Lorentz group,
$\UL(\C^2)_\R/\U(1_2)\cong\SO^+(1,3)$ .

The compact phase group $\U(1_2)$ in $\UL(\C^2)_\R$ 
is used for the representation of the 
time-space translations
in the case of particle fields (chapter 1). Those representations 
are positive unitary, but not faithful.

In general, unitary groups realizing time-space translations will be called
modality groups. They characterize the conjugations and inner products
involved and, therewith, the probability interpretation of the theory.

For  vector fields, the Lorentz group $\SO^+(1,3)$ is  embedded
into the indefinite unitary group 
$\U(1,3)$, compatible with the Lorentz 'metric'
$(-1,1,1,1)$.
The arising field types are given in the scheme
\[
\SO^+(1,3)\subnoteq\U(1,3)
\begin{array}{l}
\supnoteq \left\{\begin{array}{c}
\U(1)\o\U(3)\supnoteq \U(1)\x\SO(3)\\
\hbox{Sylvester particles}\\
(m^2>0)\end{array}\right.\\
~~\\
\supnoteq\left\{\begin{array}{c}
\U(1,1)\o\U(2)\supnoteq \U(1)\x\SO(2)\\
\hbox{Maxwell-Witt fields}\\
(m^2=0)\end{array}\right.\end{array}
\]For the Witt decomposition the indefinite Lorentz 'metric'
gives rise to  the indefinite unitary group $\U(1,1)$ 
as modality group for
the nonparticle contributions of the Maxwell-Witt fields \cite{SBH95}.

The symmetry group of a relativistic field dynamics, e.g. $\SO^+(1,3)$,
should be distinguished from the
 unitary modality group, e.g. $\U(1,3)$, which in general
is  a strictly larger group\footnote{\scriptsize 
An analogue situation is familiar from the 'dynamical symmetries',
e.g. $\U(2,2)$ for the nonrelativistic hydrogen atom containing the
symmetries $\SU(2)\x \SU(2)$ for the bound states and $\SL(\C^2)_\R$ for the
scattering states.}. 

The analogue embeddings  for
spinor fields in nondecomposable Lorentz symmetry representations involves
Majorana and Weyl particles
\[
\SL(\C^2)_\R\subnoteq\UL(\C^2)_\R
\begin{array}{l}
\supnoteq\left\{\begin{array}{c}
\U(1_2)\o \SU(2)\\
\hbox{Majorana particles}\\
(m^2>0)\end{array}\right.\\
~~\\
\supnoteq\left\{\begin{array}{c}
\U(1)\x \U(1)\\
\hbox{Weyl particles}\\
(m^2=0)\end{array}\right.\\
\end{array}
\]The stability group 
for Weyl particles is  a
$\U(1)$-circularity (polarization) 
with $\U(1)\cong\SO(2)$,
for Majorana particles spin $\SU(2)$ with $\SU(2)/\I_2\cong\SO(3)$.
The additional
$\U(1)$ group  realizes the time-space translations.

The stability group for   
Dirac particles is spin $\SU(2)$ 
and  - in addition - an internal charge group $\U(1)$ which arises because of
the twofold left-right handed Lorentz group representation 
involved 
\[
\U(1)\x \UL(\C^2)_\R\supnoteq
\left\{ \begin{array}{c}
\U(1)\x\U(1_2)\o\SU(2)\\
\hbox{Dirac particles}\\
(m^2>0)
\end{array}\right.
\]

Representations of the time-space translations in $\U(1)$,
as used for Wigner classified particles,
are
irreducible and positive unitary, but unfaithful.
Faithful, but reducible representations of $\M\cong\R^4$ are given
in the indefinite unitary modality group $\U(2,2)$ whose phase orbits
constitute the orthogonal conformal group $\SO(2,4)\cong\U(2,2)/\U(1_4)$. 
$\U(2,2)$ contains as indefinite unitary subgroup the Lorentz group
with the time-space translations 
$\U(2,2)\supnoteq\SL(\C^2)_\R\x_s\R^4$ (Poincar\'e group).

Fields with
translation representations in indefinite modality groups, 
e.g. massless gauge fields, Fa\-deev-Popov fields etc.
(chapter 2), have  no full particle analysis. 
The mathematical structures involved, especially
the connection between translations representation and metrical
structure, are sketched in chapter 3. The main problem
using fields which describe interactions without asymptotic particles
is the unitarization, i.e. the establishment of a
projection condition, compatible with the dynamics, to a state space
with a positive inner product. It is shown in chapter 4, how
the projection to translation eigenstates coincides with the
projection  of the full algebra of fields
to a subalgebra with positive inner product.
In the case of Maxwell-Witt fields, the projection 
to time-space translation eigenstates 
coincides with the familiar gauge invariance condition 
(Becchi-Rouet-Stora invariance \cite{BRS}) for
quantum gauge fields.

\chapter{PARTICLE FIELDS AND  POSITIVE MODALITY GROUP}

For a relativistic field $\bl \Phi(x|m)$ with mass $m\ge0$
which is symmetric with respect to a conjugation $*$ and 
allows an analysis of the time-space translations properties
\begin{eqn}{l}
\bl \Phi_\pm(x|m)
=\int {d^4q~  e^{ixq}\over(2\pi)^3}
{\scriptsize\pmatrix{1\cr-i\ep(q_0)\cr}}\de(m^2-q^2)~\ro e(q)
 =\bl \Phi_\pm(x|m)^*
\end{eqn}the energy-momentum reflected harmonic components   $\ro e(\pm q)$ 
are related to each other by the conjugation $*$
\begin{eqn}{l}
\bl \Phi_\pm(x|m) 
=\intqp3 {\scriptsize\pmatrix{1\cr-i\cr}}
{e^{ixq}\ro e(q)\pm e^{-ixq}\ro e(q)^*\over2 }\Big|_{q_0=\sqrt{m^2+\rvec q^2}}\\
\ro e(q)=\ro e(-q)^*\\
\end{eqn}
 
 The real 4-dimensional additive group of the time-space translations 
 $\M\cong\R^4$ is realized for particle fields
 in the real 1-dimensional compact unitary group $\U(1)$  
 with the energy-momenta $q$, $q^2=m^2$,  as eigenvalues
  \begin{eqn}{l}
 D_1(~~|q):\M\map\U(1),~~\cases{
 D_1(x|q)= e^{ixq}=D_1(-x|q)^\star\\
 \p^j|_{x=0}D_1(x|q)=iq^j\cr} \end{eqn}Because of the 
 positive definite modality group $\U(1)$ with conjugation $\star$ 
 particle fields
 have a probability interpretation.
The  time-space representation $D_1(x|q)$ 
in $\U(1)$ is irreducible and not faithful. 
 
The relation between the
$\U(1)$-conjugation $\star$ for the represented translations and the
field conjugation $*$ above has to take care of the spin properties involved.
  
 \section{Sylvester Particles}
 
 Sylvester particles will be defined as particles
 with nontrivial mass and  stability group $\SO(3)$,
 they carry integer spin representations. They are bosons.

For faithful representations of 
the  Lorentz group $\SO^+(1,3)$ 
with stability group $\SO(3)$ the defining representation  
can be exemplified by a massive vector field 
without internal charge degrees of freedom, e.g. the
free neutral weak boson field $\bl Z^k$
 of the standard model with mass $M>0$. With a rest system
determined up to space rotations, the 
time-space translations analysis for $\bl Z^k$ 
and its canonical partner $\bl G^{kj}$ read
\begin{eqn}{rl}
\bl Z(x)^k=&\intqp3\sqrt M~\La({q\over M})^k_a 
{e^{ixq}\ro U(\rvec q)^a
+ e^{-ixq}\de^{ab}\ro U(\rvec q)^\star_b\over\sqrt 2}\\
\bl G(x)^{kj}=&-i\intqp3\sqrt M~
\La({q\over M})^l_0\ep^{kj}_{lr}\La({q\over M})^r_a 
{e^{ixq}\ro U(\rvec q)^a
- e^{-ixq}\de^{ab}\ro U(\rvec q)^\star_b\over\sqrt 2}\\
&\ep^{kj}_{lr}=\de^k_l\de^j_r-\de^j_l\de^k_r
\end{eqn}

The boosts $\La({q\over M})$ with $q^2=M^2$ transmutate
from Lorentz vector fields to spinning particles, i.e. from  $\SO^+(1,3)$ to 
$\SO(3)$ representations with
three spin directions $a=1,2,3$
\begin{eqn}{l}
\La({q\over M})^k_{0,a}\cong{1\over M}
{\scriptsize\pmatrix{
q_0&\rvec q\cr
\rvec q&\bl 1_3 M+{\rvec q\ox\rvec q\over q_0+M}\cr}}
,~~\La(1,0,0,0)=\bl 1_4
\end{eqn}Those transmutators are representatives for the classes of the real
3-di\-men\-sio\-nal Sylvester manifold $\SO^+(1,3)/\SO(3)$.

The free field dynamics is illustrated by the 
classical $\SO^+(1,3)$-invariant Lagrangian
\begin{eqn}{l}
\cl L(\bl Z,\bl G)=\bl G^{jk}{\p_j\bl Z_k-\p_k\bl Z_j\over2}
-\cl I(\bl Z,\bl G)\\
\cl I(\bl Z,\bl G)=-M({\bl G^{jk}\bl G_{jk}\over4}+{\bl Z^j\bl Z_j\over2})
\end{eqn}

With the complex embedding $\SO^+(1,3)\subnoteq \U(1,3)$, the stability
group comes with a $\U(1_3)$-conjugation,
$\U(1_3)\x\SO(3)\subnoteq \U(1,3)$. The positive definite modality group $\U(1_3)$
represents the time-space 
translations. Its conjugation 
exchanges  $Z$-creation operators  $\ro U(\rvec q)^a$ 
with $Z$-annihilation operators  $\ro U(\rvec q)^\star_a$ 
\begin{eqn}{l}\left.\begin{array}{c}
\hbox{conjugation $\star$}\\
\hbox{for modality group $\U(1_3)$}
\end{array}\right\}
~~\ro U(\rvec q)^a\lrmap\de^{ab}\ro U(\rvec q)^\star_b
\end{eqn}Lorentz vector fields are symmetric with respect to the conjugation
$\star$, i.e. $\bl Z=\bl Z^\star$, $\bl G=\bl G^\star$.

The quantization and Fock-space positive inner product
\begin{eqn}{rl}
\com{\ro U(\rvec p)^\star_ a}{\ro U(\rvec q)^b}
&=\de_a^b(2\pi)^3q_0\de(\rvec q-\rvec p)\\
\angle{\acom{\ro U(\rvec p)^\star_ a}{\ro U(\rvec q)^b}}
&=\de_a^b(2\pi)^3q_0\de(\rvec q-\rvec p)
=\angle{\ro U(\rvec p)^\star_ a\ro U(\rvec q)^b}
\end{eqn}lead to the field commutators and Fock values of the anticommutators, e.g.
\begin{eqn}{rl}
{\scriptsize \pmatrix{
\com{\bl Z(y)^k}{\bl Z(x)^j}\cr
\angle{\acom{\bl Z(y)^k}{\bl Z(x)^j}}\cr}}&
=-(\eta^{kj}+{\p^k\p^j\over M^2})
{\scriptsize\pmatrix{i\bl s(x-y|M)\cr \bl c(x-y|M)\cr}}\\
&=\int{d^3q~ e^{-i(\rvec x-\rvec y)\rvec q}\over(2\pi)^3 q_0}
M\La({q\over M})^k_a \de^{ab}
{\scriptsize \pmatrix{i\sin (x_0-y_0)q_0\cr
\cos (x_0-y_0)q_0\cr} }\La({q\over M})^j_b
\end{eqn}with the quantization distribution $\bl s$ and the
expectation function $\bl c$
\begin{eqn}{rl}
{\scriptsize\pmatrix{\bl c(x|m)\cr \bl s(x|m)\cr}}&=
\int{d^4q~e^{ixq}\over(2\pi)^3}
{\scriptsize\pmatrix{
1\cr -i\ep(q_0)\cr}}\de(m^2-q^2)
=\int{d^3q~e^{-i\rvec x\rvec q}\over(2\pi)^3q_0}
{\scriptsize\pmatrix{\cos x_0q_0\cr \sin x_0q_0\cr}}\\
\end{eqn}

The modality group $\U(1_3)$,  generated by $iI(\ro U)$, 
is compatible
with the stability group $\SO(3)$, generated by $i\rvec S(\ro U)$
\begin{eqn}{c}
I(\ro U)=\intqp3 {\acom{\ro U(\rvec q)^a}{\ro U(\rvec q)^\star_ a}\over2}
=I(\ro U)^\star\\
 S(\ro U)^a=\intqp3~i\ep^{abc} 
{\acom{\ro U(\rvec q)^b}{\ro U(\rvec q)^\star_ c}\over2}
=S(\ro U)^{a\star}\\
\com{ I(\ro U)}{\rvec S(\ro U)}=0\end{eqn}

\section{Dirac Particles}

Dirac particles will be defined as particles with nontrivial mass and
stability group $\U(2)$,
they feel  half integer spin $\SU(2)$  representations 
and  a nontrivial  internal charge group $\U(1)$. They are fermions.

Faithful  representations of the phase
Lorentz group $\UL(\C^2)_\R$ with
stability group $\U(2)$
are exemplified
by massive Dirac  fields  $\bl\Psi=(\bl l,\bl r)$. They carry 
a decomposable phase Lorentz group representation with irreducible 
left and right handed Weyl contributions, illustrated by
the free electron field of the standard model with mass $m>0$. 
The time-space translations analysis for left and right handed contributions
\begin{eqn}{rl}
\bl l(x)^A&=\intqp3
\sqrt m~\la({q\over m})^A_\al
~{e^{ixq}\ro u(\rvec q)^\al
+ e^{-ixq}\ro a(\rvec q)^{\star\al}\over\sqrt 2}\\
-i\bl r(x)^{\dot A}&=-i\intqp3
\sqrt m~\hat \la({q\over m})^{\dot A}_{\al}
~{e^{ixq}\ro u(\rvec q)^\al
- e^{-ixq}\ro a(\rvec q)^{\star\al}\over\sqrt 2}\\
\bl l(x)^\star_{\dot A}&=\intqp3
\sqrt m~\la({q\over m})_{\dot A}^{\star\al} 
~{e^{ixq}\ro a(\rvec q)_\al
+ e^{-ixq}\ro u(\rvec q)^\star_\al\over\sqrt 2}\\
i\bl r(x)^\star_A&=-i\intqp3
 \sqrt m~\la({q\over m})_A^{-1\al}
~{e^{ixq}\ro a(\rvec q)_\al
- e^{-ixq}\ro u(\rvec q)^\star_\al\over\sqrt 2}
\end{eqn}involves
electron and positron  operators for 
creation  $\ro u(\rvec q)$, $\ro a(\rvec q)$ 
and annihilation $\ro u(\rvec q)^\star,\ro a(\rvec q)^\star$.

The Weyl represented boosts $\la({q\over m})$ with $q^2=m^2$ transmutate
from spinor fields to particles, i.e. from  $\SL(\C^2)_\R$ to 
$\SU(2)$ representations with two spin directions $\al=1,2$
\begin{eqn}{l}
\la({q\over m})=\sqrt{{q_0+m\over 2m}}(\bl 1_2+{\rvec\si\rvec q\over q_0+m}),~~
\hat\la({q\over m})=\la({q\over m})^{\star -1}=
\sqrt{{q_0+m\over 2m}}(\bl 1_2-{\rvec\si\rvec q\over q_0+m})\\
\la(1,0,0,0)=\bl 1_2=\hat\la(1,0,0,0)\\
\La({q\over m})^k_j
={1\over2}\tr\la({q\over m})\rho^k\la({q\over m})^\star\d\rho_j,~~
\la({q\over m})^A_\al\la({q\over m})^{\star\al}_{\dot A}
={(\rho_k)^A_{\dot A}q^k\over m}\\
\hbox{Weyl matrices: }\d\rho_k=(\bl 1_2,\rvec\si),~~\rho_k=(\bl 1_2,-\rvec\si)\\
\end{eqn}

A classical $\UL(\C^2)_\R$-invariant Lagrangian reads
\begin{eqn}{l}
\cl L(\bl l,\bl r)=i\bl l\d\rho_k\p^k\bl l^\star+i\bl r\rho_k\p^k\bl r^\star
-\cl I (\bl l,\bl r)\\
\cl I(\bl l,\bl r)=m(\bl l^A\bl r_A^\star+\bl r^{\dot A}\bl l_{\dot A}^\star)\\
\end{eqn}

The quantization connects dual pairs
\begin{eqn}{c}
\acom{\ro u(\rvec p)^\star_\al}{\ro u(\rvec q)^\be}
=\acom{\ro a(\rvec p)_\al}{\ro a(\rvec q)^{\star\be}}
=\de_\al^\be (2\pi)^3q_0\de(\rvec q-\rvec p)
\end{eqn}

The stability group conjugation
\begin{eqn}{c}
\left.
\begin{array}{c}
\hbox{conjugation $\star$}\\
\hbox{for stability group $\U(1_4)\o\SU(2_2)$}
\end{array}\right\} ~~
\cases{
\ro u(\rvec q)^\al\lrmap \de^{\al\be}\ro u(\rvec q)^\star_\be\cr
\ro a(\rvec q)^{\star\al}\lrmap \de^{\al\be}\ro a(\rvec q)_\be\cr}\\
\end{eqn}exchanges  creation and annihilation operators.

The  $\U(1_4)$ phase group, e.g. the electromagnetic charge group
for electrons and positrons is generated by $iI(\ro u,\ro a^\star)$
\begin{eqn}{rl}
I(\ro u,\ro a^\star)&=I(\ro u)+I(\ro a^\star)=\intqp3 
{\com{\ro u(\rvec q)^\al}{\ro u(\rvec q)^\star_\al }
+\com{\ro a(\rvec q)^{\star\al}}{\ro a(\rvec q)_\al }\over2}
=I(\ro u,\ro a^\star)^\star
\end{eqn}and the spin group $\SU(2_2)$ by $i\rvec S(\ro u,\ro a^\star)$
\begin{eqn}{rl}
\rvec S(\ro u,\ro a^\star)
=\rvec S(\ro u)+\rvec S(\ro a^\star)
&=\intqp3 ~\rvec\si_\al^\be
{\com{\ro u(\rvec q)^\al}{\ro u(\rvec q)^\star_\be }
+\com{\ro a(\rvec q)^{\star\al}}{\ro a(\rvec q)_\be }\over2}
=\rvec S(\ro u,\ro a^\star)^\star\\
\end{eqn}

The translations representing  group
$\U(1_2)_3$ has the generator  $iI(\ro u,\ro a)$
\begin{eqn}{c}
I(\ro u,\ro a)=I(\ro u)- I(\ro a^\star)
=\intqp3
{\com{\ro u(\rvec q)^\al}{\ro u(\rvec q)^\star_\al }
-\com{\ro a(\rvec q)^{\star\al}}{\ro a(\rvec q)_\al }\over2}
=I(\ro u,\ro a)^\star\\
\com
{I(\ro u,\ro a^\star)+\rvec S(\ro u,\ro a^\star)}
{I(\ro u,\ro a)}=0
\end{eqn}

The Fock inner product is positive with the 
stability group conjugation $\star$
\begin{eqn}{l}
\angle{\com{\ro u(\rvec p)^\star_\al}{\ro u(\rvec q)^\be}}
=\angle{\ro u(\rvec p)^\star_\al\ro u(\rvec q)^\be}
=\de_\al^\be (2\pi)^3q_0\de(\rvec q-\rvec p)\\
\angle{\com{\ro a(\rvec p)^{\star\be}}{\ro a(\rvec q)_\al}}
=\angle{\ro a(\rvec p)^{\star\be}\ro a(\rvec q)_\al}
=\de_\al^\be (2\pi)^3q_0\de(\rvec q-\rvec p)\\
\end{eqn}

Quantization and Fock state
lead to the familiar field anticommutators and Fock values of the
commutators
\begin{eqn}{l}
{\scriptsize\pmatrix{
\acom{\bl l(0)^\star}{\bl l(x)}\cr
\angle{\com{\bl l(0)^\star}{\bl l(x)}}\cr}}=\rho_k\p^k
{\scriptsize\pmatrix{\bl s(x|m)\cr -i\bl c(x|m)\cr}},~~
\acom{\bl l(0)^\star}{\bl l(\rvec x)}=\rho_0\de(\rvec x)\hbox{ etc.}\\
\end{eqn}

Spinor  fields are symmetric 
$\bl l^\dagger=\bl l$, $(i\bl r)^\dagger=i\bl r$ etc.
with respect to the indefinite conjugation
exchanging particle creation with antiparticle annihilation
\begin{eqn}{c}
\hbox{conjugation $\dagger$: }
\ro u(\rvec q)^\al\lrmap \ro a(\rvec q)^{\star\al},~~
\ro a(\rvec q)_\al\lrmap \ro u(\rvec q)^\star_\al 
\end{eqn}

\section{Weyl Particles}
Weyl particles will be defined as massless particles 
with stability group $\U(1)$ which describes  both an internal charge and 
circularity. They are fermions.

The massless limit of the $\SL(\C^2)_\R/\SU(2)$-transmutator, used for
a Dirac field, leads to the two projectors
for lightlike energy-momenta $q^2=0$, $q_0\ne0$
\begin{eqn}{l}
p_+(q)={\displaystyle\lim_{m\to0}}\sqrt{{m\over2q_0}}\la({q\over m})=
{\bl 1_2+{\rvec\si\rvec q\over|\rvec q|}\over2},~~
p_-(q)={\displaystyle\lim_{m\to0}}\sqrt{{m\over2q_0}}\hat\la({q\over m})=
{\bl 1_2-{\rvec\si\rvec q\over|\rvec q|}\over2}\\
p_+(q_0,0,0,\pm q_0)={\bl 1_2\pm\si_3\over2}=
p_-(q_0,0,0,\mp q_0)
\end{eqn}Any spacelike direction ${\rvec\si\rvec q\over|\rvec q|}$ 
can be transformed into a fixed  3rd axis $\si_3$ of 
a rest frame, determined up to  $\SO(2)$ rotations of the $(1,2)$-plane 
\begin{eqn}{l}
o({\rvec q\over q_0})~\si_3~o({\rvec q\over q_0})^\star
={\rvec\si\rvec q\over|\rvec q|},~~q_0^2=\rvec q^2>0
\end{eqn}with a 'rotation' $o({\rvec q\over q_0})\in\SU(2)$ 
as a representative of a class in
$\SO(3)/\SO(2)$ $\cong$  $\SU(2)/\U(1)_3$
\begin{eqn}{l}
o({\rvec q\over q_0})={1\over\sqrt{2q_0(q_0+q_3)}}
{\scriptsize\pmatrix{
q_0+q_3&q_1-iq_2\cr -q_1-iq_2&q_0+q_3\cr}},~~o(0,0,1)=\bl 1_2\\
p_\pm(q)=o({\rvec q\over q_0})~{\bl 1_2\pm\si_3\over2}~
o({\rvec q\over q_0})^\star=
o_\pm({\rvec q\over q_0}) o_\pm({\rvec q\over q_0})^\star\\
\hbox{with }o_\pm({\rvec q\over q_0})^A=o({\rvec q\over q_0})^A_{1,2}\\
\end{eqn}

Therewith the time-space translations analysis of a 
free massless Weyl field with a left-handed Lorentz group
representation and classical La\-gran\-gian 
$\cl L(\bl l_+)=i\bl l_+\d\rho_k\p^k\bl l_+^\star$, e.g. of the 
electron neutrino field in the
standard model  - if massless - looks as follows
\begin{eqn}{rl}
\bl l_+(x)^A&=\intqp3\sqrt{q_0}~
o_+({\rvec q\over q_0})^A
~(e^{ixq}\ro u(\rvec q)+ e^{-ixq}\ro a(\rvec q)^\star)\\
\bl l_+(x)^\star_{\dot A}&=\intqp3\sqrt{q_0}~
o_+({\rvec q\over q_0})^\star_{\dot A}
~(e^{ixq}\ro a(\rvec q)
+ e^{-ixq}\ro u(\rvec q)^\star)\\
\end{eqn}

The transmutators $o({\rvec q\over q_0})$  
represent only the real 2-dimensional manifold
$\SO(3)/\SO(2)$. For the real 5-dimensional Witt manifold
$\SO^+(1,3)/\SO(2)$  an additional
Sylvester transmutator $\SO^+(1,3)/\SO(3)$ has to be used,
irrelevant in this connection. 

With the 
massless field stability group $\U(1_2)\x \U(1)_3\subnoteq \U(2)$ there
is no $\SU(2)$-spin degree of freedom left in the particle regime.
Starting from the Dirac particles abelian stability group  
$e^{i(\al_0\bl 1_2+\al_3\si_3)}\ox\bl 1_2\in \U(1_4)\x \U(1_2)_3$ 
the  stability group $\U(1)$
for massless spinor particles
$e^{i\al_+{\bl 1_2+\si_3\over2}}\ox\bl 1_2\in \U(1_2)_+$ arises
by projection with $p_+(q)$.

The conjugation $\star$ exchanges creation with annihilation
\begin{eqn}{rl}
\left.\begin{array}{c}
\hbox{conjugation $\star$}\\
 \hbox{for stability group $\U(1_2)_+$}\end{array}\right\}
 ~~&\ro u(\rvec q)\lrmap \ro u(\rvec q)^\star,~~
\ro a(\rvec q)\lrmap \ro a(\rvec q)^\star
\end{eqn}

The stability group $\U(1)$  is generated by $iI_+(\ro u,\ro a^\star)$ 
\begin{eqn}{rl}
I_+(\ro u,\ro a^\star)&=I_+(\ro u)+ I_+(\ro a^\star)=
\intqp3 {\com{\ro u(\rvec q)}{\ro u(\rvec q)^\star }
+\com{\ro a(\rvec q)^\star}{\ro a(\rvec q) }\over2}
= I_+(\ro u,\ro a^\star)^\star\\
\end{eqn}E.g. for massless neutrinos $I_+(\ro u,\ro a^\star)$
is the fermion number or the polarization. 

The translations representing  group $\U(1)$ is generated by $iI_+(\ro u,\ro a)$  
\begin{eqn}{c}
I_+(\ro u,\ro a)=I_+(\ro u)-I_+(\ro a^\star)=\intqp3 
{\com{\ro u(\rvec q)}{\ro u(\rvec q)^\star }
-\com{\ro a(\rvec q)^\star}{\ro a(\rvec q) }\over2}
=I_+(\ro u,\ro a)^\star\\
\com{I_+(\ro u,\ro a^\star)}{I_+(\ro u,\ro a)}=0\\
\end{eqn}

The Fock product is positive with  the conjugation $\star$ - in 
 analogy to the massive case.

\section{Majorana Particles}

Majorana particles  - if they exist -
will be defined as particles with nontrivial mass  
and  stability group $\SU(2)$ for spin without an internal charge. 
They are fermions.

Since the $\SL(\C^2)_\R$
Lorentz properties  of the irreducible Weyl
contributions  $\bl l(x)^A $ and
$\bl r(x)_A^\star$ in a  Dirac field are isomorphic 
with the invariant bilinear spinor 'metric'
\begin{eqn}{l}
\la\in \SL(\C^2)_\R:~~\ep_{AB}\la^B_C\ep^{CD}=(\la^{-1})_A^D
 ,~~\ep_{AB}=-\ep_{BA}
\end{eqn}one can consider the case where
the four Dirac fields 
$(\bl l,\bl r^\star;\bl r,\bl l^\star)$ 
are built with only two irreducible left and right handed Weyl representations 
$(\bl L,\bl R)$ 
by 'crossover' identifying particles and antiparticles 
\begin{eqn}{l}
\ro a(\rvec q)^{\star\al}= i\ep^{\al\be}
\ro u(\rvec q)^{\star}_\be,~~
\ro a(\rvec q)_\al=-i\ro u(\rvec q)^\be\ep_{\be\al}
\end{eqn}

Therewith one decribes Majorana fields with the time-space translations analysis
\begin{eqn}{rl}
\bl L(x)^A&=\intqp3
\sqrt m~\la({q\over m})^A_\al
~{e^{ixq}\ro u(\rvec q)^\al
+ e^{-ixq}i\ep^{\al\be}\ro u(\rvec q)^\star_\be\over\sqrt 2}
=i\ep^{AB}\bl R(x)^\star_ B\\
\bl L(x)^\star_{\dot A}&=\intqp3
\sqrt m~\la({q\over m})_{\dot A}^{\star\al}
~{-e^{ixq}\ro u(\rvec q)^\be i\ep_{\be\al}
+ e^{-ixq}\ro u(\rvec q)^\star_\al\over\sqrt 2}
=-i\bl R(x)^{\dot B}\ep_{\dot B\dot A}\\
\end{eqn}with
classical Lagrangian, only $\SL(\C^2)_\R$ invariant
\begin{eqn}{l}
\cl L(\bl L)=i\bl L\d\rho_k\p^k\bl L^\star-\cl I(\bl L)\\
\cl I(\bl L)=im(\ep_{BA}\bl L^A\bl L^B
-\bl L^\star_{\dot A}\bl L^\star_{\dot B}\ep^{\dot B\dot A})
\end{eqn}

The two conjugations
use the two components $\al=1,2$ 
\begin{eqn}{rl}
\left.\begin{array}{c}
\hbox{conjugation $\star$}\\
\hbox{for stability group $\SU(2)$ }\end{array}\right\}
~~&
\ro u(\rvec q)^\al\lrmap \de^{\al\be}\ro u(\rvec q)^\star_\be\\
~~&\\
\hbox{conjugation $\dagger$ }
~~&
\ro u(\rvec q)^\al\lrmap i\ep^{\al\be}\ro u(\rvec q)^\star_\be\\
\end{eqn}On can write for the combinations in the
time-space translations analysis
\begin{eqn}{l}
\ro u^1=\ro u,~\ro u^2=i\ro a\then\left\{\begin{array}{l}
\ro u^\al+i\ep^{\al\be}\ro u^\star_\be\cong
\pmatrix{\ro u+\ro a^\star\cr i(\ro a-\ro u^\star)\cr}\\
i(\ro u^\al-i\ep^{\al\be}\ro u^\star_\be)\cong
\pmatrix{i(\ro u-\ro a^\star)\cr -(\ro a+\ro u^\star)\cr}\\
\end{array}\right.
\end{eqn}

The dual pair quantization and  
Fock values are analogue to the Dirac case
\begin{eqn}{l}
\acom{\ro u(\rvec p)^\star_\al}{\ro u(\rvec q)^\be}
=\de_\al^\be (2\pi)^3q_0\de(\rvec q-\rvec p)=
\angle{\com{\ro u(\rvec p)^\star_\al}{\ro u(\rvec q)^\be}}
=\angle{\ro u(\rvec p)^\star_\al\ro u(\rvec q)^\be}
\end{eqn}

The generators $i\rvec S(\ro u)$ for the 
spin group $\SU(2)$  and $iI(\ro u)$  for 
the translations realizing  group $\U(1)$ are

\begin{eqn}{c}
\rvec S(\ro u)=\intqp3~\rvec\si^\be_\al
{\com{\ro u(\rvec q)^\al}{\ro u(\rvec q)^\star_\be }\over2}
= \rvec S(\ro u)^\star\\
I(\ro u)=\intqp3 {\com{\ro u(\rvec q)^\al}{\ro u(\rvec q)^\star_\al }
\over2}= I(\ro u)^\star\\
\com{\rvec S(\ro u)}{I(\ro u)}=0
\end{eqn}

\chapter{FIELDS WITH INDEFINITE MODALITY GROUP}

Particle noninterpretable quantum fields  are used
for locally formulated interactions. They arise e.g. in 
gauge fields. The electromagnetic vector
field with its four Lorentz components
has two particle degrees of freedom
with modality group $\U(2)$, the two massless photons as
left and right polarized representations for the stability group $\SO(2)$
of the  time-space Witt-decomposition $\M\cong \T\pl\S^2\pl\S^1$.
The two  additional $\SO(2)$-trivial
lightlike degrees of freedom $\T\pl\S^1\cong\L_+\pl\L_-\cong\R^2$
without particle interpretation decribe the gauge degree of freedom and 
the Coulomb interaction. They have an
indefinite  $\U(1,1)$-modality group for the
represented time-space  translations.

Also Fadeev-Popov fields have an indefinite $\U(1,1)$-conjugation $\x$ without particle
interpretation.

A relativistic field $\bl \Phi'(x|m)={d\over dm^2}\bl \Phi(x|m)$  of mass $m\ge0$
which is conjugation $*$ symmetric and
allows an analysis of the time-space translations
\begin{eqn}{l}
\bl \Phi'_\pm(x|m)
=\int {d^4q ~e^{ixq}\over(2\pi)^3}
{\scriptsize\pmatrix{1\cr-i\ep(q_0)\cr}}\de'(m^2-q^2)~\ro e(q) 
=\bl \Phi'_\pm(x|m)^*
\end{eqn}contains  harmonic components $\ro e(q,x)$
with a 1st order polynomial dependence 
in the time-space translations
\begin{eqn}{l}
\bl \Phi'_\pm(x|m)
=\intqp3 {\scriptsize\pmatrix{1\cr-i\cr}}
{e^{ixq}\ro e(q,x)
\pm e^{-ixq}\ro e(q,x)^*\over2}\Big|_{q_0=\sqrt{m^2+\rvec q^2}}\\
\hbox{with }\ro e(q,x)=\ro e_0(q)+ix\ro e_1(q)=\ro e(-x,-q)^*\\
\end{eqn}
 
 The real 4-dimensional additive group of the time-space translations 
 $\M\cong\R^4$ is represented  in 
 the noncompact unitary conformal group $\U(2,2)$ 
 with the energy-momenta $q$, $q^2=m^2$,  as eigenvalues
  \begin{eqn}{l}
 D_2(~~|q):\M\map\U(2,2),~~\cases{
 D_2(x|q) \cases{=
 e^{ixq}{\scriptsize\pmatrix{\bl 1_2&i\rho^j x_j\cr0&\bl 1_2\cr}}
 =e^{iQ^jx_j}\\
={\scriptsize\pmatrix{\bl 1_2&\rho^j {\p\over\p q^j}\cr0&\bl 1_2\cr}}
 e^{ixq}\cr}\\
D_2(x|q)=D_2(-x|q)^\x\\
 \p^j|_{x=0}D_2(x|q)=iQ^j=i
{\scriptsize\pmatrix{q^j\bl 1_2&\rho^j\cr0&q^j\bl 1_2\cr}}\cr}
 \end{eqn}The image of the time-space translations 
 is an $\R^4$-isomorphic unitary subgroup of
 $\U(2,2)$
 as illustrated by the nondiagonalizable 'triangular' Jordan matrix 
 with the characteristic nilpotent contributions. 
 The time-space representations $D_2(x|q)$ are
 faithful and reducible, but nondecomposable.
 Because of the indefinite unitary modality group 
 such fields  have no probability interpretation in terms of particles.

\section{Maxwell-Witt Fields}

Maxwell-Witt fields\cite{SBH95} will be defined as massless 
Lorentz vector Bose fields with stability group
$\SO(2)$ for circularity (polarization). In addition to
massless particles they contain also nonparticle contributions.

The classical $\SO^+(1,3)$-invariant 
Lagrangian for a free massless vector field, e.g.
the electromagnetic field
\begin{eqn}{l}
\cl L(\bl A,\bl F,\bl G)=
\bl G\p_k\bl A^k+\bl F^{jk}{\p_j\bl A_k-\p_k\bl A_j\over2}
-\cl H(\bl A,\bl F,\bl G)\\
\cl H(\bl A,\bl F,\bl G)=-\mu{\bl F^{jk}\bl F_{jk}\over4}-
\si{\bl G^2\over2}
\end{eqn}has to include - with respect to a quantum   framework -
a canonical partner $\bl G$, called gauge fixing field, for the scalar part of
the vector field $\bl A^k$. $\mu>0$ is a mass 
(no particle mass) which - in an interacting
theory - can be related to the gauge coupling constant, $\si\ne0$ 
is called gauge fixing constant.

In the quantization distributions\cite{NAK}
\begin{eqn}{l}
{\scriptsize\pmatrix{
\com{i\bl F^{kj}(0)}{\bl A_r(x)}\cr
\com{\bl A^k(0)}{\bl G(x)}\cr
\com{\bl A^k(0)}{\bl A^j(x)}\cr}}=\int{d^4q~e^{ixq}\over(2\pi)^3}\ep(q_0)
{\scriptsize\pmatrix{
\ep^{kj}_{lr}q^l\de(q^2)\cr
q^k\de(q^2)\cr
-\mu\eta^{kj}\de(q^2)-(\mu+\si)q^kq^j\de'(q^2)\cr}}
\end{eqn}the dipole $\de'(m^2-q^2)$  
is a characteristic  feature of the nonparticle
structure
\begin{eqn}{rl}
\bl s'(x|m)&={d\over dm^2}\bl s(x|m)
=-i\int{d^4q~e^{ixq}\over(2\pi)^3}\ep(q_0)\de'(m^2-q^2)\\ &\\
&=\int {d^3q~e^{-i\rvec x\rvec q}\over(2\pi)^3 q_0}
{x_0q_0\cos x_0q_0-\sin x_0q_0\over 2q_0^2}
\end{eqn}

The time-space translations analysis of the massless vector field
has to include a transmutation  $O({\rvec q\over q_0})$ with $q^2=0$,
$q\ne0$,
from the rest frames  stability group $\SO(3)$
to $\SO(2)$ for rest frames with fixed 3rd axis
\begin{eqn}{l}
O({\rvec q\over q_0})={\scriptsize\pmatrix{
1&0&0&0\cr
0&1-{(q_1)^2\over q_0(q_0+q_3)}&-{q_1q_2\over q_0(q_0+q_3)}&{q_1\over q_0}\cr
0&-{q_1q_2\over q_0(q_0+q_3)}&1-{(q_2)^2\over q_0(q_0+q_3)}&{q_2\over q_0}\cr
0&-{q_1\over q_0}&-{q_2\over q_0}&{q_3\over q_0}\cr}}\\
O({\rvec q\over q_0})^k_j={1\over2}
\tr o({\rvec q\over q_0})\rho^k o({\rvec q\over q_0})^\star\d \rho_j,~~
O(0,0,1)=\bl 1_4\\
\end{eqn}

According to  the isomorphy
$\T\pl\S^1\cong\L_+\pl\L_-$, 
it is convenient to transform from a time-space 
Sylvester basis  with diagonal metrical tensor $\eta$
to a light-space-light Witt basis with 
'skew-diagonal' metrical tensor $\io$ 
\begin{eqn}{c}
\hbox{Sylvester: }
-\eta={\scriptsize\pmatrix{
-1&0\cr
0&\bl1_3\cr}},~~
\hbox{Witt: }
-\iota={\scriptsize\pmatrix{
0&0&1\cr
0&\bl1_2&0\cr
1&0&0\cr}}\\
\iota=w~\eta~w^T\hbox{ with }w=
{\scriptsize\pmatrix{
{1\over\sqrt2}&0&{1\over\sqrt2}\cr
0&\bl1_2&0\cr
-{1\over\sqrt2}&0&{1\over\sqrt2}\cr}}\\
\end{eqn}

The time-space translations analysis\cite{SBH95} of the massless vector field 
embeds the Lorentz group with its signature $(1,3)$ indefinite 'metric'
in an indefinite unitary group $\U(1,3)\supnoteq \SO^+(1,3)$
which determines the conjugations and modality  groups for the gauge field
\begin{eqn}{rl}
\bl A(x)^k=&\intqp3 \sqrt\mu~O({\rvec q\over q_0})^k_j w^j_{\dots}
{\scriptsize\pmatrix{
{e^{ixq}\ro B(\rvec q,x_0)+e^{-ixq}N_0\ro G(\rvec q)^\x\over\sqrt2}\cr
{e^{ixq}\ro U(\rvec q)^1+e^{-ixq}\ro U(\rvec q)^\star_ 1\over\sqrt2}\cr
{e^{ixq}\ro U(\rvec q)^2+e^{-ixq}\ro U(\rvec q)^\star _2\over\sqrt2}\cr
{e^{ixq}N_0\ro G(\rvec q)+e^{-ixq}\ro B(\rvec q,x_0)^\x\over\sqrt2}\cr}}\\
\bl G(x)=&i\intqp3\sqrt\mu~
{e^{ixq}\ro G(\rvec q)-e^{-ixq}\ro G(\rvec q)^\x\over\sqrt2}
\end{eqn}The $(1,2)$-components $\ro U(\rvec q)^{1,2}$ are
particle degrees of freedom. 
The $(0,3)$-components 
$(\ro B(\rvec q),\ro G(\rvec q))$ have a linear translation  dependence
\begin{eqn}{l}
\ro B(\rvec q,x_0)=\ro B(\rvec q)+{ix_0q_0\over M_0}\ro G(\rvec q)
\hbox{ with }\cases{
{1\over M_0}=-{\mu+\si\over\mu}\cr
N_0={3\mu+\si\over\mu}\cr}\\
\end{eqn}The characteristic terms ${ix_0q_0\over M_0}e^{ix_0q_0}$
are associated  to nondecomposable, 
but reducible  representations\cite{BOE,S89}  of the time translations
\begin{eqn}{l}
D_2(x_0|q_0)=e^{ix_0q_0}{\scriptsize\pmatrix{1& {ix_0q_0\over M_0}\cr 0&1\cr}}
=e^{ix_0q_0{\scriptsize\pmatrix{1& {1\over M_0}\cr 0&1\cr}}}
\end{eqn}as  an $\R$-isomorphic subgroup of 
$\U(1,1)$
\begin{eqn}{l}
D_2(x_0|q_0)^\x={\scriptsize\pmatrix{0&1\cr1&0\cr}}~D_2(x_0|q_0)^\star~
{\scriptsize\pmatrix{0&1\cr1&0\cr}}=D_2(-x_0|q_0)
\end{eqn}

The quantization connects dual pairs
\begin{eqn}{rl}
\hbox{for }(1,2):&~~\com{\ro U(\rvec p)^\star_\al}{\ro U(\rvec q)^\be}
=\de_\al^\be (2\pi)^3q_0\de(\rvec q-\rvec p)\\
\hbox{for }(0,3):&~~\cases{
\com{\ro G(\rvec p)^\x}{\ro B(\rvec q)}
=\com{\ro B(\rvec p)^\x}{\ro G(\rvec q)}
=(2\pi)^3q_0\de(\rvec q-\rvec p)\\
\com{\ro G(\rvec p)^\x}{\ro G(\rvec q)}=0
=\com{\ro B(\rvec p)^\x}{\ro B(\rvec q)}\cr}
\end{eqn}

The $(1,2)$-particle degrees of freedom have
a $\U(1_2)$-conjugation $\star$ whereas a 
$\U(1,1)$-conjugation $\x$  applies for the 
$(0,3)$-nonparticle degrees of freedom
\begin{eqn}{l}
\left.\begin{array}{c}
\hbox{conjugation $\star$}\\
\hbox{for $(1,2)$-modality group $\U(1_2)$}\\
\end{array}\right\} 
~~\ro U(\rvec q)^{1,2}\lrmap\ro U(\rvec q)^\star_{ 1,2}\\
~~\\
\left.\begin{array}{c}
\hbox{conjugation $\x$}\\
\hbox{for $(0,3)$-modality group $\U(1,1)$}\\
\end{array}\right\}
~~\cases{\ro G(\rvec q)\lrmap\ro G(\rvec q)^\x\cr
\ro B(\rvec q)\lrmap\ro B(\rvec q)^\x\cr}\\
\end{eqn}

The modality group as $\R$-isomorphic  
noncompact subgroup of $\U(1_2)\o\U(1,1)\subnoteq \U(1,3)$ is generated with 
\begin{eqn}{rl}
H(\ro U,\ro B,\ro G)=&\intqp3\(
{\acom{\ro U(\rvec q)^\al}{\ro U(\rvec q)^\star_\al}
+\acom{\ro B(\rvec q)}{\ro G(\rvec q)^\x}
+\acom{\ro G(\rvec q)}{\ro B(\rvec q)^\x}\over2}
+{\ro G(\rvec q)\ro G(\rvec q)^\x\over M_0}\)\\
=&I(\ro U)+H(\ro B,\ro G)= I(\ro U)^\star+H(\ro B,\ro G)^\x\\
\end{eqn}

The stability group $\SO(2)\cong \U(1)_3$ (polarization)  is
generated by $iS(\ro U)$ with the particle degrees of freedom only  
\begin{eqn}{c}
S(\ro U)=\intqp3
{\acom{\ro U(\rvec q)^1}{\ro U(\rvec q)^\star_1}
-\acom{\ro U(\rvec q)^2}{\ro U(\rvec q)^\star_2}\over2}
=S(\ro U)^\star\\
\com{H(\ro U,\ro B,\ro G)}{S(\ro U)}=0
\end{eqn}

With the $\U(1_2)$-conjugation $\star$ the Fock product for the particle
degrees of freedom is positive definite
\begin{eqn}{l}
\hbox{for }(1,2):~~
\angle{\acom{\ro U(\rvec p)^\star_\al}{\ro U(\rvec q)^\be}}
=\de_\al^\be (2\pi)^3q_0\de(\rvec q-\rvec p)=
\angle{\ro U(\rvec p)^\star_\al\ro U(\rvec q)^\be}
\end{eqn}
 
 The $\U(1,1)$-conjugation 
 $\x\cong{\scriptsize\pmatrix{0&1\cr1&0\cr}}$  
 for the  nonparticle degrees of freedom 
leads to an indefinite inner Fock-product
\begin{eqn}{l}
\hbox{for }(0,3):~~\cases{
\angle{\acom{\ro G(\rvec p)^\x}{\ro B(\rvec q)}}
=\angle{\acom{\ro B(\rvec p)^\x}{\ro G(\rvec q)}}
=(2\pi)^3q_0\de(\rvec q-\rvec p)\\
\then\angle{{\ro G(\rvec p)^\x\pm\ro B(\rvec p)^\x\over\sqrt2}
  {\ro G(\rvec q)\pm\ro B(\rvec q)\over\sqrt2}}
=\pm (2\pi)^3q_0\de(\rvec q-\rvec p)\\}
\end{eqn}

For a probability interpretation,
the indefinite metric has to be avoided for the asymptotic state space:
Fadeev-Popov fields counterbalance 
the 'negative  probabilities'. 
The requirement of gauge invariance as Becchi-Rouet-Stora invariance  
in a quantum theory projects to
a positive definite asymptotic particle subspace (chapter 4). 

\section{Fadeev-Popov Fields}

Fadeev-Popov fields will be defined as massless Lorentz scalar Fermi fields.
They have no particle contributions.

Their classical Lagrangian uses two scalar fields $\bl A_+$, $\bl U_-$ 
in a 2nd order
derivative formalism 
$\cl L(\bl A_+,\bl U_-)=i(\p^k\bl A_+)(\p_k\bl U_-)$
or, in addition, two vector fields $\bl U_+^k$,
$\bl A_-^k$ for
a 1st order formulation
\begin{eqn}{rl}
\cl L(\bl A_\pm,\bl U_\pm)&=i\bl A_+\p_k\bl U_+^k
+i\bl U_-\p_k\bl A_-^k-\cl H(\bl A_\pm,\bl U_\pm)\\
\cl H(\bl A_\pm,\bl U_\pm)&=i\mu\bl U_+^k\bl A_{-k}\\
\end{eqn}with a mass scale $\mu>0$ (no particle mass).

The quantization for the Fadeev-Popov fields with the translations analysis
\begin{eqn}{rl}
\bl A_+(x)&=\intqp3\sqrt\mu~
{e^{ixq}\ro a(\rvec q)+ e^{-ixq}\ro a(\rvec q)^\x\over\sqrt 2}\\
\bl U_-(x)&=i\intqp3\sqrt\mu~
{e^{ixq}\ro u(\rvec q)- e^{-ixq}\ro u(\rvec q)^\x\over\sqrt 2}\\
\bl U_+(x)^k&=\intqp3\sqrt\mu~\La({q\over\mu})^k_0
{e^{ixq}\ro u(\rvec q)+ e^{-ixq}\ro u(\rvec q)^\x\over\sqrt 2}\\
\bl A_-(x)^k&=i\intqp3\sqrt\mu~\La({q\over\mu})^k_0
{e^{ixq}\ro a(\rvec q)- e^{-ixq}\ro a(\rvec q)^\x\over\sqrt 2}\\
\end{eqn}connects as dual pairs 
\begin{eqn}{l}
\acom{\ro u(\rvec p)^\x}{\ro a(\rvec q)}=(2\pi)^3 q_0\de(\rvec q-\rvec p)
=\acom{\ro a(\rvec p)^\x}{\ro u(\rvec q)}
\end{eqn}

A positive $\U(1)$-conjugation $\star$ is impossible, i.e.
$\ro u^\x$ and $\ro a^\x$
cannot be identified
with $\ro a^\star$ and $\ro u^\star$ resp. With $\acom{\bl U_-}{\bl U_-}=0$
also an identification $\ro u=\ro a$ and $\ro u^\x=\ro u^\star$ cannot be used.

Therewith Faddeev-Popov fields have only the indefinite
\begin{eqn}{l}
\begin{array}{c}
\hbox{conjugation $\x$}\\
\hbox{for modality group $\U(1,1)$}\end{array}
\ro u(\rvec q)\lrmap\ro u(\rvec q)^\x,~~
\ro a(\rvec q)\lrmap\ro a(\rvec q)^\x
\end{eqn}The fields are symmetric with 
the conjugation $\x$, i.e. $\bl U_-=\bl U_-^\x$
etc.\cite{RD,KO}

The $\U(1)$ group for the time translations is generated by
$I(\ro a,\ro u)$ with
\begin{eqn}{l}
I(\ro a,\ro u)=\intqp3
{\com{\ro a(\rvec q)}{\ro u(\rvec q)^\x}
+\com{\ro u(\rvec q)}{\ro a(\rvec q)^\x}\over2}=I(\ro a,\ro u)^\x
\end{eqn}

The Fock inner product is indefinite with the $\U(1,1)$-conjugation
$\x\cong{\scriptsize\pmatrix{0&1\cr 1&0\cr}}$
\begin{eqn}{l}
\angle{   \com{\ro u(\rvec p)^\x}{\ro a(\rvec q)}  }
=\angle{   \com{\ro a(\rvec p)^\x}{\ro u(\rvec q)}  }
=(2\pi)^3 q_0\de(\rvec q-\rvec p)\\
\then\angle{ 
{\ro u(\rvec p)^\x \pm \ro a(\rvec p)^\x\over \sqrt2}   
{\ro u(\rvec q) \pm \ro a(\rvec q)\over \sqrt2}  }=\pm (2\pi)^3 q_0
\de(\rvec q-\rvec p)
\end{eqn}

\section{Heisenberg-Majorana Fields}

Heisenberg-Majorana fields will be defined as massive Lorentz spinor 
Fermi fields without particle
degrees of freedom. They can be relevant only for the description of
interactions.

Heisenberg-Majorana fields, written with left handed fields 
$\ro b^A$, $\ro g^A$, 
are ana\-ly\-sab\-le with 
the time-space translations represented in  $\U(2,2)$
\begin{eqn}{rl}
\bl b(x)^A&=\intqp3\la({q\over m})^A_\al
~{e^{ixq}\ro b(\rvec q,x)^\al+ e^{-ixq}
i\ep^{\al\be}\ro b(\rvec q,x)^\x_\be\over\sqrt 2}\\
\bl b(x)^\x_{\dot A}&=\intqp3\la({q\over m})^{\star\al}_{\dot A}
~{-e^{ixq}\ro b(\rvec q,x)^\be i\ep_{\be\al}+ e^{-ixq}
\ro b(\rvec q,x)^\x_\al\over\sqrt 2}\\
\bl g(x)^A&=\intqp3\la({q\over m})^A_\al
~{e^{ixq}\ro g(\rvec q)^\al+ e^{-ixq}
i\ep^{\al\be}\ro g(\rvec q)^\x_\be\over\sqrt 2}\\
\bl g(x)^\x_{\dot A}&=\intqp3\la({q\over m})^{\star\al}_{\dot A}
~{-e^{ixq}\ro g(\rvec q)^\be i\ep_{\be\al}+ e^{-ixq}
\ro g(\rvec q)^\x_\al\over\sqrt 2}\\
\end{eqn}The harmonic components have a linear time-space dependence
with the translations components  $x({q\over m})_k$, $k=0,1,2,3$,
written in a rest system
\begin{eqn}{l}
\ro b(\rvec q,x)^\al=\ro b(\rvec q)^\al+ix({q\over m})^\al_\be
\ro g(\rvec q)^\be\\
x({q\over m})^\al_\be=(\rho^k)^\al_\be x({q\over m})_k=
\la({q\over m})^{-1\al}_A~ x^A_{\dot A}~\hat\la({q\over m})^{\dot A}_\be\\
x({q\over m})_k=\La({q\over m})_k^{-1j}x_j,~~
x^A_{\dot A}=(\rho^k)^A_{\dot A}x_k
\end{eqn}

The quantization   of the harmonic components
connects dual pairs
\begin{eqn}{l}
\acom{\ro b(\rvec p)^\x_\al}{\ro g(\rvec q)^\be}
=\acom{\ro g(\rvec p)^\x_\al}{\ro b(\rvec q)^\be}
=\de_\al^\be(2\pi)^3q_0\de(\rvec q- \rvec p)\\
\acom{\ro g(\rvec p)^\x_\al}{\ro g(\rvec q)^\be}
=0=\acom{\ro b(\rvec p)^\x_\al}{\ro b(\rvec q)^\be}\\
\end{eqn}and leads to the field quantization
\begin{eqn}{l}
\acom{\bl b(0)^\x_{\dot A}}{\bl b(x)^A}
=-(\rho^k)^A_{\dot A} x_k~\bl s(x|m),~~
\acom{\bl g(0)^\x_{\dot A}}{\bl g(x)^A}
=0\\
\acom{\bl g(0)^\x_{\dot A}}{\bl b(x)^A}
=\acom{\bl b(0)^\x_{\dot A}}{\bl g(x)^A}
=(\rho^k)^A_{\dot A} \p_k~\bl s(x|m)\\
\end{eqn}with the dipole distribution
\begin{eqn}{rl}
{x_k\over2}\bl s(x|m)=\p_k\bl s'(x|m)
={d\over dm^2}\p_k\bl s(x|m)
&=\int{d^4q~e^{ixq}\over(2\pi)^3}\ep(q_0)q_k\de'(m^2-q^2)\\
\end{eqn}

A classical $\SL(\C^2)_\R$-invariant Lagrangian reads
\begin{eqn}{l}
\cl L(\bl b,\bl g)=
i\bl b\d\rho_k\p^k\bl g^\x
+i\bl g\d\rho_k\p^k\bl b^\x
-\cl H(\bl b,\bl g)\\
\cl H(\bl b,\bl g)
=i(\ep_{BA}\bl g^A\bl g^B
-\bl g^\x_{\dot A}\bl g^\x_{\dot B}\ep^{\dot B\dot A})
+im(\ep_{BA}\bl b^A\bl g^B
-\bl g^\x_{\dot A}\bl b^\x_{\dot B}\ep^{\dot B\dot A})
\end{eqn}

The  conjugation $\x$  for the 
time-space translations is characterized by
the unitary conformal group $\U(2,2)$
 \begin{eqn}{rl}
 \left.\begin{array}{c}
 \hbox{conjugation $\x$}\\
 \hbox{for modality group}\\
  \U(2,2)
 \end{array}\right\}
 &~~\cases{
 \ro b(\rvec q)^\al\lrmap\de^{\al\be}\ro b(\rvec q)_\be^\x\cr
 \ro g(\rvec q)^\al\lrmap\de^{\al\be}\ro g(\rvec q)_\be^\x\cr}\\
\end{eqn}

The $\R^4$-isomorphic time-space translation group is generated 
by $i Q(\ro b,\ro g)^j$
\begin{eqn}{rl}
Q(\ro b,\ro g)^j&=\intqp3\(q^j
{\com{\ro b(\rvec q)^\al}{\ro g(\rvec q)^\x_\al}
+\com{\ro g(\rvec q)^\al}{\ro b(\rvec q)^\x_\al}\over2}
+\ro g(\rvec q)^\al(\rho^j)_\al^\be\ro g(\rvec q)^\x_\be
\)\\
&=I(\ro b,\ro g)^j+ N(\ro g)^j=Q(\ro b,\ro g)^{j\x}
\end{eqn}

A compatible stability group $\U(1_4)$ is generated by $iI(\ro b,\ro g)$
\begin{eqn}{c}
I(\ro b,\ro g)=\intqp3
{\com{\ro b(\rvec q)^\al}{\ro g(\rvec q)^\x_\al}
+\com{\ro g(\rvec q)^\al}{\ro b(\rvec q)^\x_\al}\over2}
=I(\ro b,\ro g)^\x\\
\com{Q(\ro b,\ro g)^j}{I(\ro b,\ro g)}=0
\end{eqn}

The fields are symmetric under the conjugation $\dagger$, i.e.
$\bl b^\dagger=\bl b$ etc.
\begin{eqn}{l}
 \hbox{conjugation $\dagger$ }~~~~~\cases{
 \ro b(\rvec q)^\al\lrmap i\ep^{\al\be}\ro b(\rvec q)_\be^\x\cr
 \ro g(\rvec q)^\al\lrmap i\ep^{\al\be}\ro g(\rvec q)_\be^\x\cr}
 \end{eqn}

It is possible - in analogy to chapter 1 - to construct
massless Heisenberg-Weyl fields and 
massive Heisenberg-Dirac fields with an internal
charge, all with indefinite unitary $\U(2,2)$ realizations of the time-space
translations. All those fields have no particle interpretation, but may be used
for the implementation of interactions.

\chapter{MODALITY GROUPS - \\THE MATHEMATICS}

The mathematical structures of this chapter
have been used  implicitely
in the former two chapters. They are 
exhibited rather frugally in the following
- more as a glossary - and can be looked at in more detail in the 
literature\cite{ALG9,BOE,S89,S912,S922,S924}. 
 
\section{Conjugations and Unitary Groups}

A conjugation $*$  is  an antilinear isomorphism between a complex
vector space $V\cong\C^d$ and the vector space $V^T\cong\C^d$ 
of its linear forms. It defines a nondegenerate
sesquilinear form which - for a conjugation - is required to be symmetric
\begin{eqn}{rl}
\hbox{conjugation: }&
*:V\lrmap V^T,~~v,\om^*\lrmap v^*,\om\\
\hbox{dual product: }&
V^T\x V\map\C,~~(\om,u)\mape \om(u)= \dprod\om u\\
\hbox{inner product: }&
^*\sprod{~}{~}:V\x V\map \C,~~^*\sprod vu=\dprod{v^*}u=\ol{\dprod{u^*}v}
\end{eqn}In the opposite direction, each symmetric nondegenerate sesquilinear  form
of a complex vecor space $V\cong\C^d$ determines a conjugation.

With the conjugation defined between the vector space and its dual, 
a conjugation is defined
on all multilinear structures, e.g. on the $V$-endomorphisms $V\ox V^T$ by
$(v\om)^*=\om^* v^*$ etc.

Since any conjugation $*$ on $V\cong\C^d$ determines its unitary invariance group
\begin{eqn}{l}
^*\sprod{g(v)}{g(u)}=~ ^*\sprod vu\iff g\in
\U(d_+,d_-)\subnoteq\GL(\C^d),~~d=d_++d_-
\end{eqn}the $d$ different classes of conjugations are characterized by the
signatures $(d_+,d_-)$.

With a fixed conjugation of $V\cong\C^d$, 
e.g. a Euklidean $\U(d)$ conjugation $\star$,
given with a dual $(V,V^T)$-basis by $\star: e^A\lrmap\de^{AB}\d e_B$, any 
conjugation $*$ is characterizable by a linear $V$-automorphism $\star\o
*\in\GL(\C^d)$.

\section{The Indefinite Unitary Poincar\'e Group}

The  unitary conformal group 
$\U(n,n)$ and its Lie algebra $\bl u(n,n)$
for $n\ge1$  can be illustrated in a 
complex $(n+n)\x(n+n)$ matrix block
representation using  a $\U(n)$ conjugation $\star$ 
to define the $\U(n,n)$ conjugation $\x$  
with the automorphism 
$\star\o\x\cong{\scriptsize\pmatrix{0&\bl 1_n\cr\bl1_n&0\cr}}$
\begin{eqn}{rl}
&F={\scriptsize\pmatrix{a&b\cr c&d\cr}}\then F^\x
={\scriptsize\pmatrix{0&\bl 1_n\cr \bl 1_n&0\cr}}~
{\scriptsize\pmatrix{a^\star&c^\star\cr b^\star&d^\star\cr}}
~{\scriptsize\pmatrix{0&\bl 1_n\cr \bl 1_n&0\cr}}
={\scriptsize\pmatrix{d^\star&b^\star\cr c^\star&a^\star\cr}}\\
&\U(n,n)=\{G\in \GL(\C^{2n})\mid G^\x= G^{-1}\}\\
& \bl u(n,n)=\{L\mid L^\x=-L\}
\end{eqn}

$\U(n,n)$  contains a $\GL(\C^n)_\R$-isomorphic  subgroup   
 with its $\x$-an\-ti\-sym\-me\-tric Lie algebra  $\bl{gl}(\C^n_2)_\R$
 as a real $2n^2$-dimensional Lie symmetry
\begin{eqn}{rl}
&\GL(\C^n)_\R\cong\GL(\C^n_2)_\R=
\{G={\scriptsize\pmatrix{g&0\cr 0&g^{-1\star}\cr}}\}\\
&\GL(\C^n)_\R=\UL(\C^n)_\R\x\D(1_n),~~\UL(\C^n)_\R=\U(1_n)\o\SL(\C^n)_\R\cr
&\bl{gl}(\C^n)_\R\cong
\bl {gl}(\C^n_2)_\R=\{L={\scriptsize\pmatrix{l&0\cr0&-l^\star\cr}}\}\\
&\bl {gl}(\C^n)_\R=\bl u(1_n)\pl  \bl {sl}(\C^n)_\R\pl\bl d(1_n)\cong\R^{2n^2}
\end{eqn}

The real abelian Lie algebras involved are $\bl u(1_n)\cong\R$ for the phases
and $\bl d(1_n)\cong\R$ for the dilatations. The 
remaining simple Lie algebra of rank
$2(n-1)$ is the generalized Lorentz Lie algebra 
$\bl{sl}(\C^n)_\R\cong \R^{2(n^2-1)}$ with the compact $\SU(n)$-Lie algebra
\begin{eqn}{rl}
&\bl u(1_{2n})=\R{\scriptsize\pmatrix{i\bl1_n&0\cr 0&i\bl 1_n}},~~
\bl d(1_n)_3=\R{\scriptsize\pmatrix{\bl1_n&0\cr 0&-\bl 1_n}}\cr
&\bl{sl}(\C^n_2)_\R=\{{\scriptsize\pmatrix{l&0\cr0&-l^\star\cr}}\mid \tr
l=0\}\cong\R^{2(n^2-1)}\\
&\bl{su}(n_2)=\{{\scriptsize\pmatrix{il&0\cr0&il\cr}}\mid \tr
l=0,~~l=l^\star \}\cong\R^{n^2-1}
\end{eqn}A possible basis for the  Lie algebra $\bl{sl}(n)$
uses the $(n^2-1)$ generalized traceless 
Pauli, Gell-Mann etc. matrices  $\rvec\si_n=\rvec\si_n^\star$,
nontrivial for $n\ge2$ 
\begin{eqn}{l}
{\scriptsize\pmatrix{i\rvec\si_n&0\cr0&i\rvec\si_n\cr}},~~
{\scriptsize\pmatrix{\rvec\si_n&0\cr0&-\rvec\si_n\cr}}
\end{eqn}

The real Lie algebra $\bl {su}(n,n)$ contains in addition 
a translation Lie algebra $\bl t(n^2)$ as
a maximal abelian ideal
\begin{eqn} {l}
\bl t(n^2)=\{{\scriptsize\pmatrix{0&x\cr 0&0\cr}}\mid x=-x^\star\}\cong\R^{n^2}
,~~\hbox{basis: }{\scriptsize\pmatrix{0&i\bl 1_n,i\rvec \si_n\cr0&0\cr}}
\end{eqn}

The  translations 
as a semidirect factor together with the 
phase, dilatations  and  Lorentz transformations 
constitute the generalized unitary Poin\-ca\-r\'e Lie algebra 
\begin{eqn}{l}
\bl u(n,n)\supnoteq\bl {poinc}(n)=
\bl u(1_{2n})\pl\bl{sl}(\C^n_2)_\R\pl\bl d(1_n)_3\pl\bl t(n^2)\cong\R^{3n^2}\\
\hbox{with }\left\{\begin{array}{l}
\com{\bl u(1_{2n})}
{\bl u(1_{2n})\pl\bl {sl}(\C^n_2)\pl\bl d(1_n)_3\pl\bl t(n^2)}=\{0\}\cr
\com{\bl d(1_n)_3}{\bl {sl}(\C^n_2)_\R\pl\bl d(1_n)_3}=\{0\},~
\com{\bl d(1_n)_3}{\bl t(n^2)}=\bl t(n^2)\cr
 \com{\bl{sl}(\C^n_2)_\R}{\bl {sl}(\C^n_2)_\R}=\bl{sl}(\C^n_2)_\R\cr
\com{\bl{sl}(\C^n_2)_\R}{\bl t(n^2)}=\bl t(n^2)\cr
\com{\bl t(n^2)}{\bl t(n^2)}=\{0\}\end{array}\right.
\end{eqn}

\section{Unitary Poincar\'e Groups\\for Time and Time-Space}

For the generalized unitary Poincar\'e groups 
in the unitary conformal groups,
the cases $n=1$, called unitary Poincar\'e group for time 
\begin{eqn}{l}
\bl u(1,1)\supnoteq \bl{poinc}(1)=\bl u(1_2)\pl 
\bl d(1)_3\pl\bl t(1)\cong\R^3,~~\bl t(1)\cong\R
\end{eqn}and $n=2$, called unitary Poincar\'e group for Minkowski time-space 
\begin{eqn}{l}
\bl u(2,2)\supnoteq \bl{poinc}(2)=\bl u(1_4)\pl\bl d(1_2)_3
\pl \bl{sl}(\C^2_2)_\R\pl\bl t(4)\cong\R^{12},~~\bl t(4)\cong\R^4
\end{eqn}are distinguished.
Only for $n=1,2$ the defining complex $n$-dimensional  representations
of $\SL(\C^n)$ have an invariant bilinear form and, therewith, a bilinear
form on the translations - time $\R$ and time-space $\R^4$. 

For $n=1$ (time) with the trivial
group $\SL(\C^1)=\{1\}$ the bilinear form is simply the product of two
numbers which  induces a definite product 
\begin{eqn}{l}
n=1:~~\bl t(1)\ni t,s\mape ts\in\R,~~t^2\ge0\end{eqn}For $n=2$ (time-space) the 
$\SL(\C^2)$-invariant
totally antisymmetric spinor 'metric' $\ep^{AB}=-\ep^{BA}$
induces  the Lorentz 'metric' $g$ on Minkowski time-space, indefinite
with signature $(1,3)$
\begin{eqn}{l}
n=2:~~\bl t(4)\ni x,y\mape g(x,y)=g(y,x)\in\R,~ ~
\hbox{sign }g=(1,3)\end{eqn}

\section{Modality Groups}

Any  representation of the 
totally ordered additive group $(\R,+)$, called causal translations group,
in a unitary group, called modality group, 
on a complex space $V\cong\C^d$, $d=d_++d_-$ 
\begin{eqn}{l}
D:\R\map \U(d_+,d_-),~~\tau\mape D(\tau)
\end{eqn}has a conjugation $*$ which
implements the inversion of the causal group $\R$ 
\begin{eqn}{l}
D(\tau)^*=D(-\tau)\end{eqn}

Any unitary causal group representations is built by nondecomposable ones.
The nondecomposable representations 
of the causal group\cite{BOE,S89} are characterized by a scale $\mu\in\R$
and a dimension $d\in\N$. They are generated by $iH_d$ with
$H_d$ being the sum of  
the identity $\bl 1_d$ on the representation space $V\cong\C^d$ and
a power $d$ nilpotent element $N_d$
\begin{eqn}{l}
D_d(~|\mu):\R\map \U_d(\R)\subnoteq\GL(\C^d),~~\left\{\begin{array}{l}
D_d(\tau|\mu)=e^{i\tau H_d}\\
H_d=\mu\bl 1_d+  N_d\\
\hbox{for }d=1:~~ N_1=0\\
\hbox{for }d\ge2:~~\left\{\begin{array}{l}
(N_d)^{d-1}\ne0\\
( N_d)^d=0\end{array}\right.
\end{array}\right.
\end{eqn}

The modality groups of the nondecomposable representations are given by
\begin{eqn}{l}
\U_d(\R)=\left\{\begin{array}{l}
\U({d+1\over2},{d-1\over2})\hbox{ for }d=1,3,\dots\\
\U({d\over2},{d\over2})\hbox{ for }d=2,4,\dots\\
\end{array}\right.
\end{eqn}

Only the  $\U(1)$-representations 
(Fourier representations) of the causal group  $\R$
are irreducible and positive unitary,
they are not faithful
\begin{eqn}{l}
D_1(\tau|\mu)=
e^{i\tau\mu}=D_1(-\tau|\mu)^\star\in\U(1)\subnoteq\GL(\C)\end{eqn}

The lowest dimensional injective 
representations are the indefinite unitary  reducible,
but nondecomposable $d=2$ representations
\begin{eqn}{rl}
D_2(\tau|\mu)= e^{i\tau\mu}{\scriptsize\pmatrix{1&i\tau\cr 0&1\cr}}
=D_2(-\tau|\mu)^\x&= {\scriptsize\pmatrix{0&1\cr 1&0\cr}}
D_2(\tau|\mu)^\star{\scriptsize\pmatrix{0&1\cr 1&0\cr}}\\
&\in \U(1,1)\subnoteq\GL(\C^2)\end{eqn}

Their antisymmetric twofold product  
gives the irreducible representation $D_1(\tau|2\mu)$, 
their totally
symmetric products  give all nondecomposable faithful 
representations
$D_n(\tau|(n-1)\mu)$, $n=2,3,\dots$.

\section{Modality Groups for Translations}

The additive group of translations $\R^{n^2}$ has the 
irreducible, nonfaithful  Fou\-rier representations in the
positive unitary modality group $\U(1)$
\begin{eqn}{l}
D_1(~~|q):\R^{n^2}\map \U(1) ,~~D_1(x|q)=e^{i\dprod xq}
\end{eqn}characterized by a linear form $q$ ('energy-momenta') of the 
translations.

Faithful representations are possible in
the subgroups $\U(1_{2n})\x \bl T(n^2)$ of the  
unitary Poincar\'e groups with the translation group  $\bl T(n^2)=e^{\bl t(n^2)}$
\begin{eqn}{c}
D_2(~~|q):\R^{n^2}\map \U(1_{2n})\x \bl T(n^2)\subnoteq\U(n,n)\\ 
D_2(x|q)=e^{i\dprod xq}
{\scriptsize\pmatrix{\bl 1_n&i(x_0\bl 1_n+\rvec x\rvec \si_n)\cr 0&\bl 1_n\cr}}
\end{eqn}
Those representations have the
indefinite modality group $\U(n,n)$.

\section{Quantum Algebras\\and Quantum Invariants}

Any complex vector space $V\cong\C^d$ defines its quantum
algebra\cite{S922,S924}
$\bl Q_\ep(\C^{2d})$ 
of Fermi or Bose type $\ep=\pm1$ as a Clifford algebra over the
direct sum space $\bl V=V\pl V^T\cong\C^{2d}$ with the linear forms $V^T$.
The Clifford factorization of the tensor algebra $\OX\bl V$ is 
performed with the dual product, extended $\ep$-symmetrically
as bilinear form of $\bl V$,  leading to the 
characteristic Fermi and Bose (anti)commutators,
given in a dual basis $\{e^A,\d e_B\}_{A,B=1}^d$ of $(V,V^T)$
by
\begin{eqn}{l}
\hbox{in }\bl  Q_\ep(\C^{2d}),~~\ep=\pm1:~~\cases{
\com{\d e_A}{e^B}_\ep=\de_A^B\cr
\com{\d e_A}{\d e_B}_\ep=0=\com{e^A}{e^B}_\ep=0\cr}
\end{eqn}

The Lie algebra of the basic space 
endomorphisms is represented  by inner derivations of the quantum algebras.

The quantum algebra functors $\bl Q_\ep$ are exponential, i.e. the quantum
algebra of a direct sum space $V\cong V_1\pl V_2$ is isomorphic to 
the tensor product of the individual quantum algebras
\begin{eqn}{l}
\bl Q_\ep(\bl V_1\pl\bl V_2)\cong\bl Q_\ep(\bl V_1)\ox\bl Q_\ep(\bl V_2)
\end{eqn}

The quantum invariants $\C\brack I$ in a quantum algebra $\bl Q_\ep(\C^{2d})$
will be defined to be those quantum elements 
which commute with all endomorphisms of the basic vector
space $V\cong\C^d$
\begin{eqn}{l}
\C\brack I=\{a\in\bl Q_\ep(\C^{2d})\mid\com{e^A\d e_B}a=0\hbox{ for all }
A,B=1,\dots,d\}
\end{eqn}They are generated by the basic space identity or by
\begin{eqn}{l}
I={\com{e^A}{\d e_A}_{-\ep}\over 2}=e^A\d e_A-\ep{d\over2}
=-\ep\d e_Ae^A+\ep{d\over2}
\end{eqn}

Bose quantum algebras $\bl Q_-(\C^{2d})$   
have countably infinite complex dimension $\aleph_0$.
In this case the identity $I$ is transcendental 
in the quantum algebra and the
ring of invariants $\C\brack I$ is isomorphic to the complex polynomials  
in one indeterminate. 

For Fermi quantum algebras which are 
- because of  the nilquadratic basic vectors
(Pauli's principle), e.g. $e^1e^1=0$  - finite dimensional
$\bl Q_+(\C^{2d})\cong\C^{4^d}$, 
the identity $I$ is algebraic  in the quantum algebra 
\begin{eqn}{l}\hbox{in }\bl Q_+(\C^{2d}):~~
(I-{d\over2})(
I-{d\over2}+1)\cdots(I+{d\over2}-1)(I+{d\over2})=0
\end{eqn}Therefore the $I$-polynomials $\C\brack I$  have maximal degree $d$.

\section{Causal Quantum Modalities}

A complex representation of the causal group $(\R,+)$ on a complex vector space
$V\cong\C^d$ can be  canonically extended 
to the quantum algebras $\bl Q_\ep(\C^{2d})$ for the representation space. The
modality group 
$\bl U(d_+,d_-)$ of the causal group representation determines a conjugation
of the quantum algebra.

The generator $iI(\ro u)$ for a positive definite  
$\U(1)$ representation of the causal group $\R$ on the space
$V\cong\C$ is given in the quantum algebras as follows
\begin{eqn}{l}
\bl Q_\ep(\C^2)\hbox{ with conjugation $\star$ of }\U(1):~~\cases{
e=\ro u,~~\d e=\ro u^\star\\
\com{\ro u^\star}{\ro u}_\ep=1\cr
\com{\ro u^\star}{\ro u^\star}_\ep=0=\com{\ro u}{\ro u}_\ep\cr}\\
I(\ro u) =\mu{\com{e}{\d e}_{-\ep}\over 2}=
\mu{\com{\ro u}{\ro u^\star}_{-\ep}\over 2}\\
\end{eqn}

The  generator $iH(\ro b,\ro g)$ for an  indefinite
$\U(1,1)$ representation of the causal group $\R$ on the space
$V\cong\C^2$  with its semisimple and nilpotent part 
$I(\ro b,\ro g)$ and $N(\ro g)$ resp. is given in the quantum algebras
as follows
\begin{eqn}{l}
\bl Q_\ep(\C^4)\hbox{ with conjugation $\x$ of }\U(1,1):~~\cases{
e^1=\ro g,~~e^2=\ro b\\
\d e_1=\ro b^\x,~~\d e_2=\ro g^\x\\
\com{\ro g^\x}{\ro b}_\ep=1=\com{\ro b^\x}{\ro g}_\ep\cr
\com{\ro g^\x}{\ro g}_\ep=0=\com{\ro b^\x}{\ro b}_\ep\\
\hbox{etc.}\cr}\\
H(\ro b,\ro g)=
\mu{\com{\ro g}{\ro b^\x}_{-\ep}
+\com{\ro b}{\ro g^\x}_{-\ep}
\over 2}+\ro g\ro g^\x=I(\ro b,\ro g)+N(\ro g)\\
\end{eqn}

The quantized $\U(n,n)$ representations of the translations $\R^{n^2}$
in the quantum algebras $\bl Q_\ep(\C^{4n})$ have 
$n^2$ generators $iQ(\ro b,\ro g)^j$
\begin{eqn}{l}
\bl Q_\ep(\C^{4n})\hbox{ with conjugation $\x$ of }\U(n,n):~~\cases{
\ro g^A,~~\ro b^A,~~A=1,\dots,n\\
\ro b_A^\x,~~\ro g_A^\x\\
\com{\ro g_A^\x}{\ro b^B}_\ep=\de_A^B=\com{\ro b_A^\x}{\ro g^B}_\ep\cr
\com{\ro g_A^\x}{\ro g^B}_\ep=0=\com{\ro b_A^\x}{\ro b^B}_\ep\\
\hbox{etc.}\cr}\\
Q(\ro b,\ro g)^j=
q^j{\com{\ro g^A}{\ro b_A^\x}_{-\ep}
+\com{\ro b^A}{\ro g_A^\x}_{-\ep}
\over 2}+(\rho^j)_A^B\ro g^A\ro g_B^\x
=q^j I(\ro b,\ro g)+ N(\ro g)^j\\
~~~~~~~~~~~~~~~\hbox{with }\rho^j\cong(\bl 1_n,\rvec\si_n)
\end{eqn}

In spaces with
 reducible, but nondecomposable representations of the causal
group $(\R,+)$, the eigenvectors for the translations form a true subspace
of all vectors with the  action of the causal group. 

In  quantum algebras with a 
causal group representation on the basic space $V\cong\C^d$, 
the subalgebra for the eigenvectors of the translations is given
by the invariants  of the nilpotent part $N$ of the generator $H=I+N$
\begin{eqn}{l}
\bl{eigen}~\bl Q_\ep(\C^{2d})=\{a\in\bl Q_\ep(\C^{2d})\mid \com {N_d}a=0\}
\end{eqn}

Obviously for $\U(1)$-modality in the quantum algebras $\bl Q_\ep(\C^2)$,
the subalgebra for the eigenvectors is the full algebra 
\begin{eqn}{c}
d=1:~~N_1=0\then \bl{eigen}~\bl Q_\ep(\C^2)=\bl Q_\ep(\C^2)\end{eqn}

For $\U(1,1)$-modality the subalgebra for the eigenvectors is a true subalgebra
generated by the basic space  eigenvectors $\ro g,\ro g^\x$ and
the basic space identity
\begin{eqn}{rl}
d=2:~~&\{1,\ro g,\ro g^\x, I(\ro b,\ro g)={\com{\ro g}{\ro b^\x}_{-\ep}
+\com{\ro b}{\ro g^\x}_{-\ep}\over2},
\com{\ro b}{\ro g}, \com{\ro g^\x}{\ro b^\x}\}\\
&\hbox{generates }
\bl{eigen}~\bl Q_\ep(\C^4)
\end{eqn}The commutators $\com{\ro b}{\ro g}$ and $\com{\ro g^\x}{\ro b^\x}$
are nontrivial only in the Fermi quantum algebra.

For $\U(n,n)$, $n\ge2$, one obtains as generating system
\begin{eqn}{rl}
d=2n:~~&\{1,\ro g^A,\ro g_A^\x, 
I(\ro b,\ro g)={\com{\ro g^A}{\ro b_A^\x}_{-\ep}
+\com{\ro b^A}{\ro g_A^\x}_{-\ep}\over2}\mid A=1,\dots,n\}\\
&\hbox{generates }\bl{eigen}~\bl Q_\ep(\C^{4n})
\end{eqn}

\section{Fock and Heisenberg Forms\\of Quantum Algebras}

Expectation values for quantum elements
need linear quantum algebra forms\cite{S912}.
Such forms will be required to be invariant 
with respect to the adjoint action of the basic space endomorphisms, 
i.e. they can be nontrivial
only on the ring of quantum invariants $\C\brack I$, generated by the 
identity $I={\com{\d e_A}{e^A}_{-\ep}\over2}$
\begin{eqn}{rl}
\angle{~~}_d:&\bl Q_\ep(\C^{2d})\map\C,~~a\mape\angle a_d\\
&a\notin\C\brack I\then \angle a_d=0\\
\end{eqn}

Since the ring of invariants is abelian, quantum algebra forms will be
 required to be abelian thereon. Therefore they are 
completely determined by the form value 
$\angle I_d$ of the generating invariant $I$
\begin{eqn}{l}
\angle{ I^k}_d=(\angle I_d)^k,~~k=0,1,\dots\end{eqn}

In Fermi quantum algebras $\bl Q_+(\C^{2d})$
the identity $I$ is algebraic of degree $d$. Therefore  its form
value can be only one of  the zeros of the 
minimal polynomial
\begin{eqn}{l}
\hbox{in }\bl Q_+(\C^{2d}):~~
\angle I_d={d\over2},{d\over2}-1,\dots,1-{d\over2}, -{d\over2}
\end{eqn}

Since a quantum algebra $\bl Q_\ep(\C^{2d})$ 
of a vector space $V$ is isomorphic to the tensor product
of its  factors with respect to a direct sum $V\cong V_1\pl V_2$, where
$V_{1,2}$ carry nondecomposable causal group representations,
a linear form is required to be 
writable as a product form on  the corresponding quantum algebra  factors
\begin{eqn}{rl}
\bl Q_\ep(\bl V_1\pl\bl V_2)\cong
\bl Q_\ep(\bl V_1)\ox\bl Q_\ep(\bl V_2)&\then
\angle{~~}_d=\angle{~~}_{d_1}\angle{~~}_{d_2}\\
a=a_1a_2&\then\angle a_d=\angle{a_1}_{d_1}\angle{a_2}_{d_2}
\end{eqn}

Therewith the possible forms
of the 'smallest'  quantum algebras $\bl Q_\ep(\C^2)$ determine
 all quantum algebra forms, if 
 there occur only irreducible causal group representations. 
 For the irreducible representations 
 $D_1(\tau|\mu)$ of the causal group on $V\cong\C$,
 the  nonfactorizable $\bl Q_\ep(\C^2)$-forms are determined  
 by the 
possible  form values $\angle I_1$ of the identity $I$
 \begin{eqn}{rl}
\angle{~~}_1:\bl Q_\ep(\C^2)\map\C&\hbox{ determined by }\cases{
\com{\d e}e_\ep=1\cr
\angle{{\com  e{\d e}_{-\ep}\over2}}_1=\angle{ I}_1\cr}\\
\then&\angle{\d e e}_1={1-2\ep\angle{I}_1\over2}\hbox{ and }
\ep\angle{ e\d e}_1={1+2\ep\angle{I}_1\over2}
\end{eqn}For Fermi quantum algebras $\bl Q_+(\C^2)$
there are only 
two forms, determined by $\angle I_1=\mp{1\over2}$,
which trivializes one of the forms
$\angle{e\d e }_1$ or $\angle{\d ee}_1$.
This structure is taken over also for the Bose case
 \begin{eqn}{l}
\ep\angle I_1=\ep\angle{{\com e{\d e}_{-\ep}\over2}}_1=\mp{1\over2}
\then\cases{
\angle{\d e e}_1=1\hbox{ and }\ep\angle{ e\d e}_1=0\cr
\angle{\d e e}_1=0\hbox{ and }\ep\angle{ e\d e}_1=1\cr
}\cr
\hbox{$\U(1)$-conjugation: } e=\ro u,~\d e=\cases{
\ro u^\star\hbox{ for }\ep\angle I_1=-{1\over2}\cr
\ep\ro u^\star\hbox{ for }\ep\angle I_1={1\over2}\cr}\cr
\end{eqn}

With those two nonfactorizable forms 
on the quantum algebras
$\bl Q_\ep(\C^2)$  over a space with a irreducible causal group representation,
 factorizable forms
 of $\bl Q_\ep(\C^{2d})$ with signature $(d_+,d_-)$ 
 can be combined
 \begin{eqn}{l}
\hbox{Fock forms of }\bl Q_\ep(\C^{2d})\cong\OX^d\bl Q_\ep(\C^2):~~\cases{
\ep\angle I_d={d_+-d_-\over2}\\
\hbox{for }d_++d_-=d=1,2,\dots\cr}
\end{eqn}Fock forms come with the distinction of a basis 
$\{\ro u^A\}_{A=1}^d$
and a decomposition  $V\cong\PL_{A=1}^d\C \ro u^A$ 
into  irreducible
1-dimensional representation spaces for the causal group.
They can also be called  Sylvester forms or 
oscillator forms or abelian forms.

Fermi quantum algebras $\bl Q_+(\C^{2d})$ 
- in contrast to Bose quantum algebras - have  
a linear reflection between
the basic vectors $V$ and linear forms $V^T$ 
which keeps invariant the quantization, but
inverts the identity $I$
\begin{eqn}{l}
e^A\lrmap\d e_A:\cases{
\acom{\d e_A}{ e^A}\lrmap \acom{e^A}{\d e_A}\hbox{ (invariant)}\\
I={e^A\d e_A-\d e^A e_A\over2}\lrmap -I,~~\angle I_d\lrmap\angle{-I}_d
\cr}
\end{eqn}The forms of Fermi quantum algebras over vector spaces
with even dimension $d$ allow a reflection compatible trivial form value 
\begin{eqn}{l}
\hbox{on }\bl Q_+(\C^{2d}):~~\angle I_d=0\hbox{ for }d=2,4,\dots
\end{eqn}Such forms  can be combined by forms
 with $\angle I_2=0$ on $\bl Q_+(\C^4)$ over
a vector space $V\cong\C^2$ with a faithful nondecomposable representation
$D_2(\tau|\mu)$ of the causal group and an indefinite $\U(1,1)$-conjugation
\begin{eqn}{rl}
\angle{~~}_2:\bl Q_+(\C^4)\mape\C&\hbox{ determined by }\cases{
\acom{\ro g^\x}{\ro b}=1=\acom{\ro b^\x}{\ro g}\\
\angle{ \com{\ro g^\x}{\ro b}}_2=0
=\angle{ \com{\ro b^\x}{\ro g}}_2\\}\\
\then &\angle{\ro g^\x\ro b}_2=\angle{ \ro b\ro g^\x}_2
= \angle{\ro b^\x\ro g}_2=\angle{ \ro g\ro b^\x}_2={1\over2}\\
\end{eqn}

The combined forms  have signature $({d\over2},{d\over2})$ 
 \begin{eqn}{l}
\hbox{Heisenberg forms of }\bl Q_+(\C^{2d})
\cong\OX^{{d\over2}}\bl Q_+(\C^4):~~\cases{
\angle I_d=0\\\hbox{for }d=2,4,\dots\cr}
\end{eqn}Heisenberg forms come with the distinction of 
 a 'pair' basis $\{\ro g^A,\ro b^A\}_{A=1}^{{d\over2}}$
and a decomposition  $V\cong\PL_{A=1}^{{d\over2}}(\C \ro g^A+\C \ro b^A)$ 
into  nondecomposable
2-di\-men\-sio\-nal representation spaces for the causal group. They 
 can also be called Witt forms or nonabelian forms.

\section{Quantum Algebras with Inner Products}

With both a conjugation $*$ 
from the basic space  $V\cong\C^d$ induced
on a quantum algebra $\bl Q_\ep(\C^{2d})$
and a linear quantum algebra form $\angle{~~}_d$, 
which is conjugation compatible $\angle{a^*}_d=\ol{\angle a_d}$,
the quantum  algebra carries
an inner product
\begin{eqn}{l}
^*\sprod{~~}{~~}:\bl Q_\ep(\C^{2d})\x\bl Q_\ep(\C^{2d})\map\C,~~
^*\sprod ab=\angle{a^*b}_d=\ol{^*\sprod ba}
\end{eqn}

The  invariance group $\U(d_+,d_-)$ for the 
conjugation $*$ of the
basic space $V\cong\C^d$ determines the positive or indefinite
structure  of the inner product of the quantum algebra.

The factorization  of a quantum algebra
with the left ideal of the orthogonal for the inner product
(Gelfand-Naimark-Segal construction)
\begin{eqn}{l}
\bl Q_\ep(\C^{2d})^\bot=\{n\in\bl Q_\ep(\C^{2d})\mid 
\angle{a^*n}_d=~^*\sprod an=0\hbox{ for all }a\in\bl Q_\ep(\C^{2d})\}
\end{eqn}determines the vector space 
$\bl Q_\ep(\C^{2d})/\bl Q_\ep(\C^{2d})^\bot$
where the classes carry an induced nondegenerate inner  product.

\chapter{PARTICLES AND INTERACTIONS - UNITARIZATION}

Quantum fields describe both particles and interactions.
An experimenter in a laboratory uses an asymptotic  space spanned
by Wigner particle states,
which has to be interpretable with probabilities.

A free relativistic quantum  field 
$\bl \Phi(x|m)$ with mass $m\ge0$, Fermi or Bose $\ep=\pm1$,
is characterized by its 
spacelike trivial quantization distribution
(principal value integration $m^2_P$ in the energy plane)
\begin{eqn}{rl}
\com{\bl\Phi}{\bl\Phi}_\ep(x|m)&=
\com{\bl\Phi(0|m)}{\bl \Phi(x|m)}_\ep=i\bl s(x|m),~
{\p_k\over m}\bl s(x|m),\dots\\
&=0\hbox{ for } x^2<0\\
i\bl s(x|m)&=\int{d^4q~e^{ixq}\over(2\pi)^3}\ep(q_0)\de(m^2-q^2)
{i\ep(x_0)\over\pi}\int{d^4q~e^{ixq}\over(2\pi)^3}{1\over m^2_P-q^2}\cr
\end{eqn}and its expectation  function for the 'opposite' commutator, 
which may be supported time-, light- and spacelike
\begin{eqn}{rl}
\angle{\com{\bl\Phi}{\bl\Phi}_{-\ep}}(x|m)&=
\angle{\com{\bl\Phi(0|m)}{\bl \Phi(x|m)}_{-\ep}}
= \bl c(x|m),~-{i\p_k\over m}\bl c(x|m),\dots\cr
\bl c(x|m)&=\int{d^4q~e^{ixq}\over(2\pi)^3}\de(m^2-q^2)\\
\end{eqn}The expectation function  - not the 
causally supported quantization distribution -
relies on the metrical structure of the quantum fields with respect
to the inner product induced by both a linear quantum algebra form and a
conjugation (chapter 3),
connected with the time-space translations representations. 

The sum of causally ordered
quantization distribution and expectation function is the Feynman propagator
\begin{eqn}{rl}
\angle{\cl T\bl\Phi\bl\Phi}(x|-im)
&= -\ep(x_0)\com{\bl\Phi}{\bl \Phi}_\ep(x|m)
+\angle{\com{\bl\Phi}{\bl \Phi}_{-\ep}}(x|m)\\
&= \bl e(x|-im),~-{i\p_k\over m}\bl e(x|-im),\dots\\
\end{eqn}together with the  conjugated distribution given as follows
\begin{eqn}{rl}
\bl e(x|\pm im)&=\pm i\ep(x_0)\bl s(x|m)+\bl c(x|m)
=\ol{\bl e(x|\mp im)}\\
&=\int{d^4q~e^{ixq}\over(2\pi)^3}2\th(\pm x_0q_0)\de(m^2-q^2)\\
&=\pm{i\over\pi}\int{d^4q~e^{ixq}\over(2\pi)^3}{1\over m^2\pm io-q^2}
=\int{d^3 q~e^{-i\rvec x\rvec q}\over(2\pi)^3 q_0}e^{\pm i|x_0|q_0}\\
\end{eqn}The quantization distribution $\ep(x_0)\bl s(x|m)$
with $\ep(x_0q_0)$ compensates in- and outgoing structures
(no spacelike contributions), in the
expectation function $\bl c(x|m)$  there occur both in- and outcoming structures.
The combinations with the  sign functions 
${1\pm\ep(x_0q_0)\over2}=\th(\pm x_0q_0)$,
relating to each other
the causal structures  of  time-space translations and energies, allow
either nontrivial  in- or nontrivial outgoing states.

The time integrals of the Feynman distributions 
exhibit via the Yukawa potential the interaction structure of the 
relativistic quantum fields. They
involve only the quantization
distribution $\bl s(x|m)$ and are independent of the inner product structure
\begin{eqn}{l}
\mp i\int dx_0 ~\bl e(x|\pm im)=\int d|x_0|~\bl s(x|m)=
{ e^{-|\rvec x|m}\over2\pi|\rvec x|}
\end{eqn}Here time and energy integration have been interchanged.

The space integral of the Feynman distributions gives a
causally ordered time representation  
\begin{eqn}{l}
\int d^3 x~ \bl e(x|\pm im)={e^{\pm i|x_0| m}\over m}\\
\int d^3 x~\bl e(\rvec x|\pm im)=\int d^3 x~\bl c(\rvec x|m)={1\over m}
\end{eqn}Here space and momentum integration have been interchanged.
For time $x_0=0$ only the 
inner product dependent expectation function $\bl c(\rvec x|m)$ contributes
nontrivially.

A quantum algebra for fields with an indefinite modality group
$\U(n,n)$ carries an indefinite inner product 
(chapter 3) which  leads 
via the expectation function  to 'negative probabilities'. The
dangerous  
quantum algebra elements with 'negative norm'
are relevant for a local formulation of
relativistic interactions,  e.g. the Coulomb interactions (section 2.1),
Since they have no particle interpretation and 
have to be avoided as in- and outgoing
states, they should contribute only with their 
interaction describing quantization distributions. 

The nilpotent part in the representation of the 
time-space translations provides
a projection to cut out a subalgebra of 
time-space translations eigenvectors (particles) which
carry a positive definite  inner product and gives rise to
the asymptotic state space.

\section{Unitarity for Particle Fields}
The 
realization of the probabilistic structure for 
relativistic fields with a complete
particle interpretation (chapter 1) is simple: Such fields
represent the time-space translations in the group $\U(1)$ or 
- more exactly in $\U(1_d)$ for $d$ degrees of freedom - 
generated by
\begin{eqn}{l}
\com{\ro u^\star_\be}{\ro u^\al}_\ep=\de_\be^\al,~~
I(\ro u)=\SUM_{\al=1}^d I(\ro u^\al)\hbox{ with }I(\ro u^\al)
={\com{\ro u^\al}{\ro u^\star_\al}_{-\ep}\over2}
\end{eqn}

For fields with momentum dependent harmonic components $\ro u(\rvec q)$ 
one has to include a 
sum with $\intqp3$.  The local stability group, e.g.
spin $\SU(2)$ and $\SO(3)$ or circularity (polarization) $\U(1)$ and $\SO(2)$,
has to be compatible with the modality group $\U(1_d)$.

The quantum algebra $\bl Q_\ep(\C^{2d})$ 
for the harmonic components $\ro u^\al,\ro u^\star_\al$ is the product of  $d$ 
individual quantum algebras  $\bl Q_\ep(\C^2)$ for each $\al$. They  carry 
via the Fock form $\angle {~~}_1$ and the $\U(1)$-conjugation $\star$
a positive definite inner product (sections 3.8, 3.9), 
e.g. shown in an orthogonal 
$\bl Q_\ep(\C^2)$-basis
$\{\ro u^k\ro u^{\star l}\mid k,l=0,1,\dots\}$
(for Fermi algebras only $k,l=0,1$)
\begin{eqn}{l}\hbox{for }\bl Q_\ep(\C^2):~~
\angle{I(\ro u)}_1=-{\ep\over2}\then
\cases{\angle{(\ro u^\star\ro u)^k}_1=1\\
\angle{\ro u^{k\star}\ro u^l}_1=k!\de_{kl}\\
^\star\sprod{\ro u^k\ro u^{\star m}}{\ro u^l\ro u^{\star n}}
=k!\de_{m0}\de_{n0}\de_{kl}\cr}
\end{eqn}

The asymptotic particle Fock space can be spanned by the classes of the
norm nontrivial vectors $\{\ro u^k\mid k=0,1,\dots\}$.

\section{Unitarization for Gauge Fields}

The dangerous indefinite structures for  Maxwell-Witt fields 
$\bl A(x)^k$ arise
because of the representation of the translation group 
for the $(0,3)$-degrees of freedom in the indefinite
unitary group $\U(1,1)$ - with the symbols of section 2.1
\begin{eqn}{l}
\hbox{in }\bl Q_-(\C^4): \com{\ro G^\x}{\ro B}=1=\com{\ro B^\x}{\ro G}\\
H(\ro B,\ro G)={\acom{\ro B}{\ro G^\x}+\acom{\ro G}{\ro B^\x}\over2}
+{\ro G\ro G^\x\over M_0}= I(\ro B,\ro G)+ N(\ro G)\\
\end{eqn}In contrast to $\ro G,\ro G^\x$, the vectors
$\ro B, \ro B^\x$ are no  eigenvectors of the time translations.
They have to be avoided in the asymptotic particle space.

The Fock form $\angle{~~}_2$ 
with the $\U(1,1)$-conjugation $\x$ 
gives an indefinite inner product
$^\x\sprod ab=\angle{a^\x b}_2$ of the Bose quantum algebra
$\bl Q_-(\C^4)$
\begin{eqn}{l}
\angle{I(\ro B,\ro G)}_2=1 \then\cases{
\angle{\ro B^\x\ro G}_2=1=\angle{\ro G^\x\ro B}_2\cr
\angle{\ro G^\x\ro G}=0=\angle{\ro B^\x\ro B}\cr
\angle{{\ro G^\x\pm\ro B^\x\over\sqrt2}
{\ro G\pm\ro B\over\sqrt2}}_2
=~^\x\sprod{{\ro G\pm\ro B\over\sqrt2}}
{{\ro G\pm\ro B\over\sqrt2}}=\pm 1\cr
}
\end{eqn}

Asymptotic help comes from the Fadeev-Popov fields  
(section 2.2) which
have a 'twin' structure with respect to the
$(0,3)$-components of the Maxwell-Witt fields
\begin{eqn}{l}
\hbox{in }\bl Q_+(\C^4): \acom{\ro a^\x}{\ro u}=1=\acom{\ro u^\x}{\ro a}\\
H(\ro a,\ro u)={\com{\ro u}{\ro a^\x}+\com{\ro a}{\ro u^\x}\over2}
+{\ro u\ro u^\x\over N_0}= I(\ro a,\ro u)+N(\ro u)
 \end{eqn}They have an indefinite Fock inner product too 
\begin{eqn}{l}
\angle{I(\ro a,\ro u)}_2=-1\then\cases{ 
\angle{\ro a^\x\ro u}_2=1=\angle{\ro u^\x\ro a}_2\cr
\angle{\ro u^\x\ro u}=0=\angle{\ro a^\x\ro a}\cr
\angle{{\ro u^\x\pm\ro a^\x\over\sqrt2}
{\ro u\pm\ro a\over\sqrt2}}_2
=~^\x\sprod{{\ro u\pm\ro a\over\sqrt2}}
{{\ro u\pm\ro a\over\sqrt2}}=\pm 1\cr
}
\end{eqn}

The generator for the translation group 
representation in $\U(1,1)\x \U(1,1)$
\begin{eqn}{l}
H(\ro B,\ro G,\ro a,\ro u)=H(\ro B,\ro G)+ H(\ro a,\ro u)
\end{eqn}is invariant under the Becchi-Rouet-Stora 
transformation\cite{BRS}
 which replaces the classical gauge transformation. The
BRS-transformation is effected by a nilquadratic  
Fermi element in the product quantum algebra
$\bl Q_-(\C^4)\ox\bl Q_+(\C^4)$ which is compatible with the
translations action\cite{S911,S913}
\begin{eqn}{l}
N(\ro G,\ro u)=i(\ro G\ro u^\x-\ro u\ro G^\x),~~
N(\ro G,\ro u)^2=0,~~
\com{H(\ro B,\ro G,\ro a,\ro u)}{N(\ro G,\ro u)}=0
\end{eqn}The BRS-charge
$N(\ro G,\ro u)$ acts by the hybrid bracket 
$\bra N a$ on the quantum elements,
i.e. with a commutator on Bose and an anticommutator on Fermi elements.

Only cooperating translation 
eigenfields $\bl G(x)$ (gauge fixing Bose field) and 
$\bl U(x)_+^k$ (Fadeev-Popov Fermi field) 
can be combined to a nilpotent Lo\-rentz vector current $\bl N(x)^j$ 
in a field theory\cite{KO} 
\begin{eqn}{c}
 N(\ro G,\ro u)=\int d^3 x~\bl N(\rvec x)^0,~~\bl N(x)^j=
 \bl G(x)\bl U(x)_+^j
\end{eqn}not the gauge fixing or the Fadeev-Popov field alone -
they give a Lorentz scalar $\bl G\bl G$ or a tensor $\bl U_+^k\bl U_+^l$.

The subalgebra of the BRS-invariants ('gauge invariants')
can be generated and spanned by translation eigenvectors only
\begin{eqn}{l}
\bl{eigen }~\bl Q_{+,-}(\C^8)=
\{a\in\bl Q_-(\C^4)\ox\bl Q_+(\C^4)\mid \bra {N(\ro G,\ro u)}a=0\}\\
\hbox{generated by }\{1,\ro G,\ro G^\x,\ro u,\ro u^\x,
I(\ro B,\ro G)+I(\ro a,\ro u)\}\\
\angle{ I(\ro B,\ro G)+I(\ro a,\ro u)}_2=0
\end{eqn}With respect to the Fock form,
this subalgebra carries a positive semidefinite inner product.
After factorization with the orthogonal of the Fock form on the
BRS-invariant subalgebra (GNS-construction), 
there remains a trivial 'c-number' complex
1-dimensional asymptotic  
vector space whose basis can be represented by the quantum
algebra unit $1$. 

Nevertheless the time-space translations representation
in the modality group  $\U(1,1)$ is relevant for the interactions
as illustrated by the ordered time integral of the quantization
distribution $\bl s(x|0)$ 
which has nontrivial contributions from both particle  and nonparticle
degrees of freedom (Coulomb potential)
\begin{eqn}{l}
i\int dx_0\ep(x_0)\com{\bl A(0)^k}{\bl A(x)^j}
 =\eta^{kj}{\mu\over 2\pi|\rvec x|}
\end{eqn}

If an 'incoming' 
particle state $s$, as a translation eigenstate  
BRS-invariant $\bra Ns=0$, e.g. with photons $\ro U^{1,2}$ and other 
particle representations $\ro u^\al$ with  modality group $\U(1)$,
undergoes a time-space 
development with the translation group generator $H$, the resulting 'outgoing'
 state $\com Hs$ remains BRS-invariant,
$\bra N{\com Hs}=0$ since $\com HN=0$.

The condition of gauge invariance, adequately implemented as
BRS-in\-va\-ri\-an\-ce for quantum fields, merges with the condition 
to have only translation eigenstates in the asymptotic state space.

\section{Unitarization for\\ Heisenberg-Majorana Fields}

Heisenberg-Majorana fields  realize 
faithfully  spa\-ce-ti\-me translations with $i Q(\ro b,\ro g)^j$
in the indefinite modality group $\U(2,2)$ -
formulated  in the notation of section 2.3
without the momenta dependence  $\ro b(\rvec q)$ etc.
\begin{eqn}{l}
\hbox{in }\bl Q_+(\C^8):~~
\acom{\ro b^\x_\al}{\ro g^\be}
=\acom{\ro g^\x_\al}{\ro b^\be}=\de_\al^\be\\
Q(\ro b,\ro g)^j=q^j
{\com{\ro b^\al}{\ro g^\x_\al}+\com{\ro g^\al}{\ro b^\x_\al}\over2}+
\ro g^\al(\rho^j)_\al^\be\ro g^\x_\be=q^j I(\ro b,\ro g)+N(\ro g)^j\\
\end{eqn}$\ro g^\al,\ro g_\al^\x$ are translation eigenvectors in contrast
to $\ro b^\al$, $\ro b_\al^\x$.

The subalgebra with all 
time-space translations eigenvectors is characterized by a trivial
action for the nilpotent part of the time-space translations representation
\begin{eqn}{l}
\bl{eigen}~\bl Q_+(\C^8)=\{a\in\bl Q_+(\C^8)\mid \com{N(\ro g)^j} a=0\}\\
\hbox{generated by }
\{1,\ro g^\al,\ro g_\al^\x, I(\ro b,\ro g)\}
\end{eqn}Obviously, the nilpotent part (nilcharge) 
is compatible with the generators of the time-space translations
\begin{eqn}{l}
\com{Q(\ro b,\ro g)^j}{N(\ro g)^k}=0
\end{eqn} 

In the full field theoretical
formulation one has the nilcurrent $\bl N(x)^j$ for the
nilcharge $N(\ro g)^j$
\begin{eqn}{c}
N(\ro g)^j=\int d^3 x~\bl N(\rvec x)^j,~~
\bl N(x)^j=\bl g(x)^A(\rho^j)_A^{\dot A}\bl g(x)^\x_{\dot A}
\end{eqn}

The appropriate quantum algebra form for the modality group $\U(2,2)$ 
is the indefinite Heisenberg form  (section 3.8)
\begin{eqn}{l}
\angle I_4=0\then\cases{ 
\angle{\com{\ro b^\x_\al}{\ro g^\be}}_4
=\angle{\com{\ro g^\x_\al}{\ro b^\be}}_4=0\\
\angle{\ro b^\x_\al\ro g^\be}_4
=\angle{\ro g^\x_\al\ro b^\be}_4=\\
\angle{\ro g^\be\ro b^\x_\al}_4
=\angle{\ro b^\be\ro g^\x_\al}_4={1\over2}\de^\be_\al\cr}
\end{eqn}With respect to the indefinite
inner product there survives only a trivial complex 1-dimensional
asymptotic state space, spanned by the quantum algebra unit $1$
(section 3.7).

A vanishing form for the generator of the translations
leads to a trivial expectation function for the Heisenberg-Majorana
fields
\begin{eqn}{l}
\angle{\com{\bl b(0)^\x_{\dot A}}{\bl b(x)^A}}=0,~~
\angle{\com{\bl g(0)^\x_{\dot A}}{\bl b(x)^A}}=0,~~
\angle{\com{\bl g(0)^\x_{\dot A}}{\bl g(x)^A}}=0\\
\end{eqn}Without spacelike contributions in the Feynman propagators,
there are no in- and outgoing particle states
\cite{HEI} or - formulated
otherwise
- the  in- and outgoing states   
compensate each other. Such a compensation is familiar from the 'twin'
structure for the $(0,3)$-gauge field contributions and the 
two Fadeev-Popov degrees of freedom (section 4.2). 

Nevertheless, Heisenberg-Majorana fields can induce 
nontrivial interactions via their causally supported 
quantization distributions, e.g. seen in the exponential potential
\begin{eqn}{rl}
\int dx_0\ep(x_0)~\acom{\bl b(0)^\x_{\dot A}}{\bl b(x)^A}
&=-2(\rho^a)_{\dot A}^A\p_a\int d|x_0|~\bl s'(x|m)\\
2\int d|x_0|~\bl s'(x|m)&=-{e^{-|\rvec x|m}\over\pi m}\\
\end{eqn}

\end{document}